\documentclass[aps,prd,onecolumn,groupedaddress,showpacs,nofootinbib,amssymb]{revtex4-2}
\usepackage[dvips]{graphicx}
\usepackage{amssymb}
\usepackage{amsmath}
\usepackage{graphicx,,color}
\usepackage{amsfonts}
\usepackage{bm}
\usepackage{cancel}
\usepackage{comment}
\usepackage{floatflt}
\usepackage{slashed}

\newcommand\be{\begin{equation}}
\newcommand\ee{\end{equation}}

\newcommand\e{\mathrm{e}}

\allowdisplaybreaks[4]
\begin{document}

\title{Propagation of Gravitational Waves in Einstein-Gauss-Bonnet Gravity for Cosmological and Spherically Symmetric Spacetimes}
\author{Shin'ichi~Nojiri$^{1,2,3}$}\email{nojiri@gravity.phys.nagoya-u.ac.jp}
\author{S.D.~Odintsov$^{3,4}$}\email{odintsov@ice.csic.es}
\author{V.K. Oikonomou,$^{5}$}\email{voikonomou@gapps.auth.gr; v.k.oikonomou1979@gmail.com}

\affiliation{ $^{1)}$ Department of Physics, Nagoya University,
Nagoya 464-8602, Japan \\
$^{2)}$ Kobayashi-Maskawa Institute for the Origin of Particles
and the Universe, Nagoya University, Nagoya 464-8602, Japan \\
$^{3)}$ Institute of Space Sciences (ICE, CSIC) C. Can Magrans
s/n, 08193 Barcelona, Spain \\
$^{4)}$ ICREA, Passeig Luis Companys, 23, 08010 Barcelona, Spain\\
$^{5)}$ Department of Physics, Aristotle University of
Thessaloniki, Thessaloniki 54124, Greece}

\begin{abstract}
In this work, we examine the propagation of gravitational waves in
cosmological and astrophysical spacetimes in the context of
Einstein-Gauss-Bonnet gravity, in view of the GW170817 event. The
perspective we approach the problem is to obtain a theory which
can produce a gravitational wave speed that is equal to that of
light in the vacuum, or at least the speed can be compatible with
the constraints imposed by the GW170817 event. As we show, in the
context of Einstein-Gauss-Bonnet gravity, the propagation speed of
gravity waves in cosmological spacetimes can be compatible with
the GW170817 event, and we reconstruct some viable models.
However, the propagation of gravity waves in spherically symmetric
spacetimes violates the GW170817 constraints, thus it is
impossible for the gravitational wave that propagates in a
spherically symmetric spacetime to have a propagating speed which
is equal to that of light in the vacuum. The same conclusion
applies to the Einstein-Gauss-Bonnet theory with two scalars. We
discuss the possible implications of our results on spherically
symmetric spacetimes.
\end{abstract}


\maketitle

\section{Introduction}\label{Sec1}

During the next decade, the focus of modern theoretical physics
and cosmology will be entirely on stage four Cosmic Microwave
Background (CMB) \cite{CMB-S4:2016ple, SimonsObservatory:2019qwx}
and gravitational wave experiments \cite{Hild:2010id,
Baker:2019nia, Smith:2019wny, Crowder:2005nr, Smith:2016jqs,
Seto:2001qf, Kawamura:2020pcg, Bull:2018lat,
LISACosmologyWorkingGroup:2022jok}. Both these experiments will
shed light on the fundamental question of whether inflation ever
occurred. Even recently, the inflationary scenario has been
considerably constrained, since the NANOGrav 2023 stochastic
gravitational wave background observation \cite{NANOGrav:2023gor}
requires a strongly blue-tilted inflationary era in order to
consistently describe the signal \cite{sunnynew}, if standard
post-inflationary cosmological scenarios occurred. Thus theories
that can yield a mild blue-tilted era can be important
phenomenologically. In this line of research, string-inspired
theories of gravity like Einstein-Gauss-Bonnet theories may play
an important phenomenological role and for an important stream of
reviews and research articles on this topic, see, for example,
\cite{Nojiri:2010wj, Nojiri:2017ncd, Hwang:2005hb, Nojiri:2006je,
Nojiri:2005vv, Satoh:2007gn, Yi:2018gse, Guo:2009uk,
Jiang:2013gza, Kanti:2015pda, vandeBruck:2017voa, Kanti:1998jd,
Pozdeeva:2020apf, Pozdeeva:2021iwc, Koh:2014bka,
Bayarsaikhan:2020jww, Chervon:2019sey, DeLaurentis:2015fea,
Nozari:2017rta, Odintsov:2018zhw, Kawai:1998ab, Yi:2018dhl,
vandeBruck:2016xvt, Kleihaus:2019rbg, Bakopoulos:2019tvc,
Maeda:2011zn, Bakopoulos:2020dfg, Ai:2020peo, Odintsov:2020xji,
Oikonomou:2020sij, Odintsov:2020zkl, Odintsov:2020mkz,
Venikoudis:2021irr, Kong:2021qiu, Easther:1996yd,
Antoniadis:1993jc, Antoniadis:1990uu, Kanti:1995vq, Kanti:1997br,
Easson:2020mpq, Rashidi:2020wwg, Odintsov:2023aaw,
Odintsov:2023lbb,Oikonomou:2022xoq,Odintsov:2023weg,
Nojiri:2023jtf,TerenteDiaz:2023kgc,Kawai:2023nqs,Kawai:2021edk,Kawai:2017kqt,Choudhury:2023kam}
and references therein. The Einstein-Gauss-Bonnet theories were
severely restricted by the GW170817 neutron star merger event
\cite{TheLIGOScientific:2017qsa, Monitor:2017mdv, GBM:2017lvd}
which indicated that the speed of the gravitational waves should
nearly coincide with that of the light in the vacuum. This
GW170817 event imposed a severe constraint on the form of the
scalar coupling factor of the scalar field on the Gauss-Bonnet
invariant, which is a function often denoted by $\xi(\phi)$, and
several scenarios were developed for the construction of a
GW170817-compatible Einstein-Gauss-Bonnet theory
\cite{Odintsov:2020xji, Odintsov:2020sqy, Oikonomou:2021kql,
Oikonomou:2022ksx}. The constraint is, however, valid only in the
Friedmann-Lema\^{i}tre-Robertson-Walker (FLRW) Universe background
and it has been shown that any constraint cannot be satisfied
around the static and spherically symmetric spacetime, which
includes black holes, stellar objects, and wormholes
\cite{Nojiri:2023jtf}.

In this paper, we again show that it is impossible to obtain a
model where the propagating speed of the gravitational wave
coincides with that of light in spherically symmetric spacetimes.
We also consider the gravitational wave speed in cosmological
spacetime and a scenario in which the Gauss-Bonnet coupling
function $\xi(\phi)$ asymptotically approaches a constant in the
late Universe when the speed of the gravitational wave has been
observed, although the Gauss-Bonnet coupling may play important
roles in the early Universe. This cosmological scenario is a
possible description of the early Universe, and it is diverse
compared to some previous approaches \cite{Odintsov:2020xji,
Odintsov:2020sqy, Oikonomou:2021kql, Oikonomou:2022ksx}. Although
from a theoretical point of view, this scenario basically
indicates that for some reason the graviton changes its mass at
late times, becoming entirely massless, it is nevertheless a
possibility that might be examined. We construct a realistic model
of Einstein-Gauss-Bonnet gravity, where the coupling function
rapidly goes to a constant in the late Universe. The model could
describe the whole evolutionary history of the Universe, including
inflation, reheating, the late accelerating expansion, and so on.
We firstly consider a standard Einstein-Gauss-Bonnet gravity
coupled with one scalar field $\phi$, and we clarify the condition
that the propagating speed of the gravitational waves is equal to
that of light and we show that the matter fluids do not affect the
propagation speed. Since the propagating speed of the
gravitational waves cannot coincide with that of light in a
non-trivial spherically symmetric background, as shown in
\cite{Nojiri:2023jtf}, we consider the scenario that the
Gauss-Bonnet coupling function $\xi(\phi)$ goes to a constant in
the late Universe and the propagating speed of the gravitational
wave approaches that of light. We construct a more realistic model
by using this scenario. The model describes both the inflationary
era in the early Universe and the accelerating expansion of the
present Universe, without introducing the parameters without
hierarchy. We also estimate the speed of the gravitational wave in
the epochs of inflation and at the end of the inflationary. We
also discuss the reheating era and estimate the temperature and
the propagating speed of the gravitational waves in this epoch. In
Section~\ref{SecIIInew}, we investigate the propagation of the
gravitational wave in the background of a spherically symmetric
spacetime. First, we consider the spherically symmetric and also
time-dependent spacetime but also static and spherically symmetric
spacetimes, and we show that the condition is not satisfied in the
non-trivial spacetime including black holes, stellar objects,
wormholes, etc. After that, we estimate the propagating speed of
the gravitational wave inside the stellar objects. We give a
general constraint when we require that the observational results
in the GW170817 event should be applied inside the stellar
objects. Also, we consider a model of the Einstein-Gauss-Bonnet
gravity coupled with two scalars, in a spherically symmetric but
time-dependent background. We show, however, that it is impossible
to obtain a model where the propagating speed of the gravitational
wave coincides with that of light. The last section is devoted to
the summary and discussion.

\section{Scalar--Einstein-Gauss-Bonnet gravity}\label{Sec1B}

First, we consider the Einstein-Gauss-Bonnet gravity, with the action of the theory being given by,\footnote{
In \cite{Anson:2019uto}, a model similar to (\ref{I8one}) has been studied and claimed that a scalar--Gauss-Bonnet coupling
could induce tachyonic instabilities in perturbations during accelerated epochs.
In the model, although the cosmological term with a cosmological constant appears in the paper, the model does not include the potential
for the scalar field as $V (\phi)$ in (\ref{I8one}).
In the model  .(\ref{I8one}), the effective potential is given by
\[
V_\mathrm{eff}\left(\phi\right) =  V(\phi) + \xi(\phi) \mathcal{G} =  V(\phi) + 24 \left( H^3 H' + H^4 \right) \xi(\phi) \, .
\]
In the last equality, we have assumed the FLRW Universe
(\ref{FRW}), $H\equiv \frac{1}{a} \frac{da}{dt}$. Therefore as
long as $V_\mathrm{eff}''\left(\phi\right) >0$, which corresponds
to the effective mass of the scalar field, the tachyon does not
appear. The condition $V_\mathrm{eff}''\left(\phi\right) >0$ could
correspond to the slow-roll condition and we can construct the
model to satisfy the condition $V_\mathrm{eff}''\left(\phi\right)
>0$. }
\begin{align}
\label{I8one}
S_{\phi\chi} = \int d^4 x \sqrt{-g} & \left\{ \frac{R}{2\kappa^2}
 - \frac{1}{2} \partial_\mu \phi \partial^\mu \phi - V (\phi) - \xi(\phi) \mathcal{G} + \mathcal{L}_\mathrm{matter} \right\}\, ,
\end{align}
where $V(\phi)$ is the potential for $\phi$, $\xi(\phi)$ is also a function of $\phi$, and finally
$\mathcal{L}_\mathrm{matter}$ denotes the Lagrangian density of the matter perfect fluids.
Furthermore, $\mathcal{G}$ is the Gauss-Bonnet invariant defined by,
\begin{align}
\label{eq:GB}
\mathcal{G} = R^2-4R_{\alpha \beta}R^{\alpha \beta}+R_{\alpha \beta \rho \sigma}R^{\alpha \beta \rho \sigma}\, .
\end{align}
By the variation of the action (\ref{I8}) with respect to the metric $g_{\mu\nu}$, we obtain,
\begin{align}
\label{gb4bD4one}
0= &\, \frac{1}{2\kappa^2}\left(- R_{\mu\nu} + \frac{1}{2} g_{\mu\nu} R\right)
+ \frac{1}{2} g_{\mu\nu} \left\{
 - \frac{1}{2} \partial_\rho \phi \partial^\rho \phi - V (\phi)\right\}
+ \frac{1}{2} \partial_\mu \phi \partial_\nu \phi \nonumber \\
&\, - 2 \left( \nabla_\mu \nabla_\nu \xi(\phi,\chi)\right)R
+ 2 g_{\mu\nu} \left( \nabla^2 \xi(\phi,\chi)\right)R
+ 4 \left( \nabla_\rho \nabla_\mu \xi(\phi,\chi)\right)R_\nu^{\ \rho}
+ 4 \left( \nabla_\rho \nabla_\nu \xi(\phi,\chi)\right)R_\mu^{\ \rho} \nonumber \\
&\, - 4 \left( \nabla^2 \xi(\phi,\chi) \right)R_{\mu\nu}
 - 4g_{\mu\nu} \left( \nabla_\rho \nabla_\sigma \xi(\phi,\chi) \right) R^{\rho\sigma}
+ 4 \left(\nabla^\rho \nabla^\sigma \xi(\phi,\chi) \right) R_{\mu\rho\nu\sigma}
+ \frac{1}{2} T_{\mathrm{matter}\, \mu\nu} \, ,
\end{align}
and the field equation for the scalar field is obtained by
varying the action with respect to $\phi$, and it is given by
\begin{align}
\label{I10one}
0 =&\, \nabla^\mu \partial_\mu \phi - V' - \xi' \mathcal{G} \, .
\end{align}
In (\ref{gb4bD4one}), $T_{\mathrm{matter}\, \mu\nu}$ is the
energy-momentum tensor of the perfect matter fluids, which obeys
the continuity equation.

\subsection{Gravitational Waves in Einstein-Gauss-Bonnet Gravity}\label{Sec1BsubA}

In this subsection, we consider the condition that the propagating
speed of the gravitational waves is equal to that of light in
vacuum, and we show that matter is irrelevant to the speed as long
as the matter minimally couples with gravity.

For the general variation of the metric,
\begin{align}
\label{variation1}
g_{\mu\nu}\to g_{\mu\nu} + h_{\mu\nu}\, ,
\end{align}
we have the following formulae in leading order in terms of
$h_{\mu\nu}$,
\begin{align}
\label{variation2}
\delta\Gamma^\kappa_{\mu\nu} =&\, \frac{1}{2}g^{\kappa\lambda}\left(
\nabla_\mu h_{\nu\lambda} + \nabla_\nu h_{\mu\lambda} - \nabla_\lambda h_{\mu\nu}
\right)\, ,\nonumber \\
\delta R^\mu_{\ \nu\lambda\sigma}=&\, \nabla_\lambda \delta\Gamma^\mu_{\sigma\nu}
 - \nabla_\sigma \delta \Gamma^\mu_{\lambda\nu}\, ,\nonumber \\
\delta R_{\mu\nu\lambda\sigma}=&\, \frac{1}{2}\left[\nabla_\lambda \nabla_\nu h_{\sigma\mu}
 - \nabla_\lambda \nabla_\mu h_{\sigma\nu}
 - \nabla_\sigma \nabla_\nu h_{\lambda\mu}
 + \nabla_\sigma \nabla_\mu h_{\lambda\nu}
+ h_{\mu\rho} R^\rho_{\ \nu\lambda\sigma}
 - h_{\nu\rho} R^\rho_{\ \mu\lambda\sigma} \right] \, ,\nonumber \\
\delta R_{\mu\nu} =& \frac{1}{2}\left[\nabla^\rho\left(\nabla_\mu h_{\nu\rho}
+ \nabla_\nu h_{\mu\rho}\right) - \nabla^2 h_{\mu\nu}
 - \nabla_\mu \nabla_\nu \left(g^{\rho\lambda}h_{\rho\lambda}\right)\right] \nonumber \\
=&\, \frac{1}{2}\left[\nabla_\mu\nabla^\rho h_{\nu\rho}
+ \nabla_\nu \nabla^\rho h_{\mu\rho} - \nabla^2 h_{\mu\nu}
 - \nabla_\mu \nabla_\nu \left(g^{\rho\lambda}h_{\rho\lambda}\right)
 - 2R^{\lambda\ \rho}_{\ \nu\ \mu}h_{\lambda\rho}
+ R^\rho_{\ \mu}h_{\rho\nu} + R^\rho_{\ \mu}h_{\rho\nu} \right]\, ,\nonumber \\
\delta R =&\, -h_{\mu\nu} R^{\mu\nu} + \nabla^\mu \nabla^\nu h_{\mu\nu}
 - \nabla^2 \left(g^{\mu\nu}h_{\mu\nu}\right)\, .
\end{align}
Then the variation of (\ref{gb4bD4one}) is given by,
\begin{align}
\label{gb4bD4B}
0=&\, \left[ \frac{1}{4\kappa^2} R + \frac{1}{2} \left\{
 - \frac{1}{2} \partial_\rho \phi \partial^\rho \phi - V \right\}
 - 4 \left( \nabla_{\rho} \nabla_\sigma \xi \right) R^{\rho\sigma} \right] h_{\mu\nu} \nonumber \\
&\, + \bigg[ - \frac{1}{4} g_{\mu\nu} \partial^\tau \phi \partial^\eta \phi
 - 2 g_{\mu\nu} \left( \nabla^\tau \nabla^\eta \xi\right)R
 - 4 \left( \nabla^\tau \nabla_\mu \xi\right)R_\nu^{\ \eta} - 4 \left( \nabla^\tau \nabla_\nu \xi\right)R_\mu^{\ \eta}
+ 4 \left( \nabla^\tau \nabla^\eta \xi \right)R_{\mu\nu} \nonumber \\
&\, + 4g_{\mu\nu} \left( \nabla^\tau \nabla_\sigma \xi \right) R^{\eta\sigma}
+ 4g_{\mu\nu} \left( \nabla_\rho \nabla^\tau \xi \right) R^{\rho\eta}
 - 4 \left(\nabla^\tau \nabla^\sigma \xi \right) R_{\mu\ \, \nu\sigma}^{\ \, \eta}
 - 4 \left(\nabla^\rho \nabla^\tau \xi \right) R_{\mu\rho\nu}^{\ \ \ \ \eta}
\bigg] h_{\tau\eta} \nonumber \\
&\, + \frac{1}{2}\left\{ 2 \delta_\mu^{\ \eta} \delta_\nu^{\ \zeta} \left( \nabla_\kappa \xi \right)R
 - 2 g_{\mu\nu} g^{\eta\zeta} \left( \nabla_\kappa \xi \right)R
 - 4 \delta_\rho^{\ \eta} \delta_\mu^{\ \zeta} \left( \nabla_\kappa \xi \right)R_\nu^{\ \rho}
 - 4 \delta_\rho^{\ \eta} \delta_\nu^{\ \zeta} \left( \nabla_\kappa \xi \right)R_\mu^{\ \rho} \right. \nonumber \\
&\, \left. + 4 g^{\eta\zeta} \left( \nabla_\kappa \xi \right) R_{\mu\nu}
+ 4g_{\mu\nu} \delta_\rho^{\ \eta} \delta_\sigma^{\ \zeta} \left( \nabla_\kappa \xi \right) R^{\rho\sigma}
 - 4 g^{\rho\eta} g^{\sigma\zeta} \left( \nabla_\kappa \xi \right) R_{\mu\rho\nu\sigma}
\right\} g^{\kappa\lambda}\left( \nabla_\eta h_{\zeta\lambda} + \nabla_\zeta h_{\eta\lambda} - \nabla_\lambda h_{\eta\zeta} \right) \nonumber \\
&\, + \left\{ \frac{1}{4\kappa^2} g_{\mu\nu} - 2 \left( \nabla_\mu \nabla_\nu \xi\right) + 2 g_{\mu\nu} \left( \nabla^2\xi\right) \right\}
\left\{ -h_{\mu\nu} R^{\mu\nu} + \nabla^\mu \nabla^\nu h_{\mu\nu} - \nabla^2 \left(g^{\mu\nu}h_{\mu\nu}\right) \right\} \nonumber \\
&\, + \frac{1}{2}\left\{ \left( - \frac{1}{2\kappa^2} - 4 \nabla^2 \xi \right) \delta^\tau_{\ \mu} \delta^\eta_{\ \nu}
+ 4 \left( \nabla_\rho \nabla_\mu \xi\right) \delta^\eta_{\ \nu} g^{\rho\tau}
+ 4 \left( \nabla_\rho \nabla_\nu \xi\right) \delta^\tau_{\ \mu} g^{\rho\eta}
 - 4g_{\mu\nu} \nabla^\tau \nabla^\eta \xi \right\} \nonumber \\
&\, \qquad \times \left\{\nabla_\tau\nabla^\phi h_{\eta\phi}
+ \nabla_\eta \nabla^\phi h_{\tau\phi} - \nabla^2 h_{\tau\eta}
 - \nabla_\tau \nabla_\eta \left(g^{\phi\lambda}h_{\phi\lambda}\right)
 - 2R^{\lambda\ \phi}_{\ \eta\ \tau}h_{\lambda\phi}
+ R^\phi_{\ \tau}h_{\phi\eta} + R^\phi_{\ \tau}h_{\phi\eta} \right\} \nonumber \\
&\, + 2 \left(\nabla^\rho \nabla^\sigma \xi \right)
\left\{ \nabla_\nu \nabla_\rho h_{\sigma\mu}
 - \nabla_\nu \nabla_\mu h_{\sigma\rho}
 - \nabla_\sigma \nabla_\rho h_{\nu\mu}
 + \nabla_\sigma \nabla_\mu h_{\nu\rho}
+ h_{\mu\phi} R^\phi_{\ \rho\nu\sigma}
 - h_{\rho\phi} R^\phi_{\ \mu\nu\sigma} \right\} \nonumber \\
&\, + \frac{1}{2}\frac{\partial T_{\mathrm{matter}\,
\mu\nu}}{\partial g_{\tau\eta}}h_{\tau\eta} \, ,
\end{align}
where we have assumed that the perfect matter fluids are minimally
coupled with gravity. We now choose a condition to fix the gauge
as follows,
\begin{align}
\label{gfc}
0=\nabla^\mu h_{\mu\nu}\, .
\end{align}
Since we are interested in the massless spin-two tensor mode, we
also impose the traceless condition,
\begin{align}
\label{ce}
0=g^{\mu\nu} h_{\mu\nu} \, .
\end{align}
We do not consider the perturbation of the scalar mode in the metric like the trace part, which may couple with the scalar field $\phi$
(see \cite{Dalang:2020eaj} for example)
because we are now interested in the massless and spin-two mode, which corresponds to the usual gravitational wave.
As long as we consider the leading order of the perturbation, the massless spin-two mode does not mix with the scalar mode,
which is a massive spin-zero mode although the second-order perturbation of the scalar field plays the role of the source of the gravitational wave.
The traceless condition (\ref{ce}) makes the massless spin-two mode decouple with the massive spin-zero mode.


Then Eq.~(\ref{gb4bD4B}) is reduced as follows,
\begin{align}
\label{gb4bD4BoneGW}
0=&\, \left[ \frac{1}{4\kappa^2} R + \frac{1}{2} \left\{
 - \frac{1}{2} \partial_\rho \phi \partial^\rho \phi - V \right\}
 - 4 \left( \nabla_\rho \nabla_\sigma \xi \right) R^{\rho\sigma} \right] h_{\mu\nu} \nonumber \\
&\, + \bigg[ \frac{1}{4} g_{\mu\nu} \left\{
 - A \partial^\tau \phi \partial^\eta \phi
 - B \left(\partial^\tau \phi \partial^\eta \chi + \partial^\eta \phi \partial^\tau \chi \right)
 - C \partial^\tau \chi \partial^\eta \chi \right\} \nonumber \\
&\, - 2 g_{\mu\nu} \left( \nabla^\tau \nabla^\eta \xi\right)R
 - 4 \left( \nabla^\tau \nabla_\mu \xi\right)R_\nu^{\ \eta} - 4 \left( \nabla^\tau \nabla_\nu \xi\right)R_\mu^{\ \eta}
+ 4 \left( \nabla^\tau \nabla^\eta \xi \right)R_{\mu\nu} \nonumber \\
&\, + 4g_{\mu\nu} \left( \nabla^\tau \nabla_\sigma \xi \right) R^{\eta\sigma}
+ 4g_{\mu\nu} \left( \nabla_{\rho} \nabla^\tau \xi \right) R^{\rho\eta}
 - 4 \left(\nabla^\tau \nabla^\sigma \xi \right) R_{\mu\ \, \nu\sigma}^{\ \, \eta}
 - 4 \left(\nabla^\rho \nabla^\tau \xi \right) R_{\mu\rho\nu}^{\ \ \ \ \eta}
\bigg\} h_{\tau\eta} \nonumber \\
&\, + \frac{1}{2}\left\{ 2 \delta_\mu^{\ \eta} \delta_\nu^{\ \zeta} \left( \nabla_\kappa \xi \right)R
 - 4 \delta_\rho^{\ \eta} \delta_\mu^{\ \zeta} \left( \nabla_\kappa \xi \right)R_\nu^{\ \rho}
 - 4 \delta_\rho^{\ \eta} \delta_\nu^{\ \zeta} \left( \nabla_\kappa \xi \right)R_\mu^{\ \rho} \right. \nonumber \\
&\, \left. + 4g_{\mu\nu} \delta_\rho^{\ \eta} \delta_\sigma^{\ \zeta} \left( \nabla_\kappa \xi \right) R^{\rho\sigma}
 - 4 g^{\rho\eta} g^{\sigma\zeta} \left( \nabla_\kappa \xi \right) R_{\mu\rho\nu\sigma}
\right\} g^{\kappa\lambda}\left( \nabla_\eta h_{\zeta\lambda} + \nabla_\zeta h_{\eta\lambda} - \nabla_\lambda h_{\eta\zeta} \right) \nonumber \\
&\, - \left\{ \frac{1}{4\kappa^2} g_{\mu\nu} - 2 \left( \nabla_\mu \nabla_\nu \xi \right) + 2 g_{\mu\nu} \left( \nabla^2\xi \right) \right\}
R^{\mu\nu} h_{\mu\nu} \nonumber \\
&\, + \frac{1}{2}\left\{ \left( - \frac{1}{2\kappa^2} - 4 \nabla^2 \xi \right) \delta^\tau_{\ \mu} \delta^\eta_{\ \nu}
+ 4 \left( \nabla_\rho \nabla_\mu \xi \right) \delta^\eta_{\ \nu} g^{\rho\tau}
+ 4 \left( \nabla_\rho \nabla_\nu \xi \right) \delta^\tau_{\ \mu} g^{\rho\eta}
 - 4g_{\mu\nu} \nabla^\tau \nabla^\eta \xi \right\} \nonumber \\
&\, \qquad \times \left\{ - \nabla^2 h_{\tau\eta} - 2R^{\lambda\ \phi}_{\ \eta\ \tau}h_{\lambda\phi}
+ R^\phi_{\ \tau}h_{\phi\eta} + R^\phi_{\ \tau}h_{\phi\eta} \right\} \nonumber \\
&\, + 2 \left(\nabla^\rho \nabla^\sigma \xi \right)
\left\{ \nabla_\nu \nabla_\rho h_{\sigma\mu}
 - \nabla_\nu \nabla_\mu h_{\sigma\rho}
 - \nabla_\sigma \nabla_\rho h_{\nu\mu}
 + \nabla_\sigma \nabla_\mu h_{\nu\rho}
+ h_{\mu\phi} R^\phi_{\ \rho\nu\sigma}
 - h_{\rho\phi} R^\phi_{\ \mu\nu\sigma} \right\} \nonumber \\
&\, + \frac{1}{2}\frac{\partial T_{\mathrm{matter}\, \mu\nu}}{\partial g_{\tau\eta}}h_{\tau\eta} \, .
\end{align}
The observation of GW170817 gives the constraint on the propagating speed $c_\mathrm{GW}$ of the
gravitational wave as follows,
\begin{align}
\label{GWp9} \left| \frac{{c_\mathrm{GW}}^2}{c^2} - 1 \right| < 6
\times 10^{-15}\, ,
\end{align}
where $c$ denotes the speed of light. In order to investigate if
the propagating speed $c_\mathrm{GW}$ of the gravitational wave
$h_{\mu\nu}$ could be different from that of the light $c$, we
only need to check the parts including the second derivatives of
$h_{\mu\nu}$,
\begin{align}
\label{second}
I_{\mu\nu} \equiv&\, I^{(1)}_{\mu\nu} + I^{(2)}_{\mu\nu} \, , \nonumber \\
I^{(1)}_{\mu\nu} \equiv&\, \frac{1}{2}\left\{ \left( - \frac{1}{2\kappa^2} - 4 \nabla^2 \xi \right) \delta^\tau_{\ \mu} \delta^\eta_{\ \nu}
+ 4 \left( \nabla_\rho \nabla_\mu \xi\right) \delta^\eta_{\ \nu} g^{\rho\tau}
+ 4 \left( \nabla_\rho \nabla_\nu \xi\right) \delta^\tau_{\ \mu} g^{\rho\eta}
 - 4g_{\mu\nu} \nabla^\tau \nabla^\eta \xi \right\} \nabla^2 h_{\tau\eta} \, , \nonumber \\
I^{(2)}_{\mu\nu} \equiv &\, 2 \left(\nabla^\rho \nabla^\sigma \xi \right)
\left\{ \nabla_\nu \nabla_\rho h_{\sigma\mu}
 - \nabla_\nu \nabla_\mu h_{\sigma\rho}
 - \nabla_\sigma \nabla_\rho h_{\nu\mu}
 + \nabla_\sigma \nabla_\mu h_{\nu\rho} \right\} \, .
\end{align}
Since we are assuming that the matter fluids minimally couple with
gravity, any contribution of the matter fluids does not couple with
any derivative of $h_{\mu\nu}$, and the contribution does not appear in $I_{\mu\nu}$.
In other words, matter is not relevant to the propagating speed of the gravitational wave.
We should note that $I^{(1)}_{\mu\nu}$ does not change the speed of the gravitational wave from the speed of light.
On the other hand, $I^{(2)}_{\mu\nu}$ changes the speed of the gravitational wave
from that of the light in general, which may violate the constraint (\ref{GWp9}).
If $\nabla_\mu \nabla^\nu \xi$ is proportional to the metric $g_{\mu\nu}$,
\begin{align}
\label{condition}
\nabla_\mu \nabla^\nu \xi = \frac{1}{4}g_{\mu\nu} \nabla^2 \xi \, ,
\end{align}
$I^{(2)}_{\mu\nu}$ does not change the speed of the gravitational
wave from that of light.
We should note that $\xi$ is a function specifying the model. Eq.~(\ref{condition})
is a condition for models so that the propagating speed of the gravitational wave coincides with that of light.

As long as we consider the FLRW Universe, we can find the solution
of Eq.~(\ref{condition}) as explicitly given in
(\ref{FLRWsolsol}). As shown in \cite{Nojiri:2023jtf}, the
condition Eq.~(\ref{condition}) cannot be satisfied in more
general background like non-trivial spherically symmetric
spacetime. Then instead of considering non-trivial solution of
Eq.~(\ref{condition}), we will consider the scenario where $\xi$
goes to a constant or vanishes in the late Universe as the
gravitational waves have been detected and $I^{(2)}_{ij}$ can be
neglected in the late Universe.

\subsection{Method for Reconstructing Realistic Models of Early and Late-time Cosmic Expansion}\label{Sec7}

In this subsection, we will focus on the scenario that the
Gauss-Bonnet coupling function $\xi(\phi)$ goes to a constant in
the late-time Universe and also that the propagating speed of the
gravitational wave approaches that of light.
In this line of research, we construct a realistic model of cosmic expansion.
Both the inflationary era in the early Universe and the accelerating
expansion of the present Universe can be described in a unified way in this model,
without introducing various parameters with different scales.
The speed of the gravitational wave in the
epochs of the inflation and at the end of the inflation are also
estimated.

We consider the Friedmann-Lema\^{i}tre-Robertson-Walker (FLRW)
Universe with a flat spatial section, the line element of which
is given by,
\begin{align}
\label{FRW}
ds^2= -dt^2 + a(t)^2\sum_{i=1,2,3} \left(dx^i\right)^2 \, .
\end{align}
Here $t$ is the cosmic time, and $a(t)$ denotes the scale factor.
We often use $H= \frac{\dot a}{a}$ which is the Hubble rate.
Now we have,
\begin{align}
\label{E2}
& \Gamma^t_{ij}= a^2 H \delta_{ij}\, ,\quad \Gamma^i_{jt}=\Gamma^i_{tj}=H\delta^i_{\ j}\, , \nonumber \\
& R_{itjt}= -\left(\dot H + H^2\right)a^2h_{ij}\, ,\quad
R_{ijkl}= a^4 H^2 \left(\delta_{ik} \delta_{lj} - \delta_{il} \delta_{kj}\right)\, ,\nonumber \\
& R_{tt}=-3\left(\dot H + H^2\right)\, ,\quad R_{ij}= a^2
\left(\dot H + 3H^2\right) \delta_{ij}\, ,\quad R= 6\dot H + 12
H^2\, , \quad \mbox{other components}=0\, ,
\end{align}
therefore if $\xi$ only depends on the cosmic time $t$, we obtain,
\begin{align}
\label{xis}
\nabla_t \nabla_t \xi= \ddot \xi \, , \quad
\nabla_i \nabla_j \xi = - a^2 H \delta_{ij} \dot \xi \, , \quad \nabla_t \nabla_i \xi = \nabla_i \nabla_t \xi =0 \, , \quad
\nabla^2 \xi = - \ddot \xi - 3 H \dot \xi \, .
\end{align}
Then Eq.~(\ref{condition}) becomes a second-order ordinary differential equation with respect to the cosmological time $t$ as follows,
\begin{align}
\label{FLRWcond}
\ddot \xi = H \dot \xi \, ,
\end{align}
whose solution is
\begin{align}
\label{FLRWsolsol}
\dot \xi = \xi_0 + \xi_1 \int dt a(t)\, .
\end{align}
Because the differential equation (\ref{FLRWcond}) is
second-order, the solution includes two constants of the
integration $\xi_0$ and $\xi_1$. In the case $\xi_1=0$, $\xi$
becomes a constant and therefore the Gauss-Bonnet term becomes a
total derivative term and the term does not contribute to any
equation. We should note that the condition $\xi_1=0$ is not an
initial condition or something else but the condition $\xi_1=0$ is
the condition defining the model. Instead of choosing the
condition $\xi_1=0$, which makes the Gauss-Bonnet term trivial, if
we choose the condition $\xi_1\neq 0$, $\xi$ is given by a
function of the cosmological time $t$. Furthermore, if we use the
relation $H=\frac{dN}{dt}$, the function $\xi$ can be expressed as
a function of the $e$-foldings $N$. By using the relation between
the scalar field $\phi$ and the $e$-foldings $N$ in
(\ref{SEGB14}), which appears later, we can determine $\xi$ as a
function of the scalar field $\phi$, which specifies the model.

We should note, however, that the propagating speed of the
gravitational wave cannot be equal to that of light in the
non-trivial spherically symmetric background, as it was shown in
\cite{Nojiri:2023jtf}.

Since the propagating speed of the gravitational wave cannot be
equal to the speed of light near black holes or stellar objects,
we consider the scenario that $I^{(2)}_{ij}$ can be neglected in the late Universe.
This requires that $\xi$ goes to a constant or vanishes in the late Universe,
which is a special solution of (\ref{condition}) corresponding to $\xi_1=0$ in (\ref{FLRWsolsol}).
In this solution,
the Gauss-Bonnet term in the action (\ref{I8one}) becomes a total
derivative and does not give any contribution to the expansion of
the Universe, although the term may become necessary in the early Universe.
If this scenario is realized, then the theory reduces to
the scalar-tensor theory in the late Universe.

The equations corresponding to the FLRW equations have the following forms, which are given by Eq.~(\ref{gb4bD4one}),
\begin{align}
\label{SEGB3}
0=&\, - \frac{3}{\kappa^2}H^2 + \frac{1}{2}{\dot\phi}^2 + V(\phi)
+ 24 H^3 \frac{d \xi(\phi(t))}{dt}\, ,\nonumber \\
0=&\, \frac{1}{\kappa^2}\left(2\dot H + 3 H^2 \right) + \frac{1}{2}{\dot\phi}^2 - V(\phi)
 - 8H^2 \frac{d^2 \xi(\phi(t))}{dt^2}
 - 16H \dot H \frac{d\xi(\phi(t))}{dt} - 16 H^3 \frac{d \xi(\phi(t))}{dt} \nonumber \\
0=&\, \ddot \phi + 3H\dot \phi + V'(\phi) + \xi'(\phi) \mathcal{G}\, .
\end{align}
The third equation in (\ref{SEGB3}) can be obtained by combining the first and second equations and therefore we forgot the third equation
in the following
By using the $e$-foldings number $N$ defined by $a=a_0\e^N$ instead of the cosmic time $t$, we now rewrite (\ref{SEGB3}) as follows,
\begin{align}
\label{SEGB3N}
0=&\, - \frac{3}{\kappa^2}H^2 + \frac{1}{2}H^2 \phi'(N)^2 + V(\phi) + 24 H^4 \frac{d \xi(\phi(N))}{dN}\, ,\nonumber \\
0=&\, \frac{1}{\kappa^2}\left(2H \frac{dH}{dN} + 3 H^2 \right) + \frac{1}{2}H^2 \phi'(N)^2 - V(\phi)
 - 8H^4 \frac{d^2 \xi(\phi(t))}{dN^2}
 - 24 H^3 \frac{dH}{dN} \frac{d\xi(\phi(N))}{dN} - 16 H^4 \frac{d \xi(\phi(t))}{dN} \, .
\end{align}
We should note $\frac{d}{dt}=H \frac{d}{dN}$ and therefore $\frac{d^2}{dt^2}= H^2 \frac{d^2}{dN^2} + \frac{dH}{dN}\frac{d}{dN}$.
By deleting $V(\phi)$ in (\ref{SEGB3N}), we obtain
\begin{align}
\label{GBeq1}
0=&\, \frac{2}{\kappa^2} H'(N) + H(N) \phi'(N)^2
 - 8 H(N)^3 \frac{d^2 \xi(\phi(t))}{dN^2}
 - 24 H(N)^2 \frac{dH}{dN} \frac{d\xi(\phi(N))}{dN} +8 H(N)^3 \frac{d \xi(\phi(t))}{dN} \nonumber \\
=&\, \frac{2}{\kappa^2} H'(N) + H(N) \phi'(N)^2
 - 8 \e^{N} \frac{d}{dN} \left( \e^{-N}H(N)^3 \frac{d \xi(\phi(t))}{dN} \right)\, ,
\end{align}
which can be integrated with respect to $\xi(N)$ and we obtain,
\begin{align}
\label{SEGB10}
\xi(\phi(N))=&\, \frac{1}{8}\int^N dN_1 \frac{\e^{N_1}}{H(N_1)^3} \int^{N_1} \frac{dN_2}{\e^{N_2}}
\left(\frac{2}{\kappa^2}H' (N_2) + H(N_2) {\phi'(N_2)}^2 \right)\, .
\end{align}
By substituting Eq.~(\ref{SEGB10}) into the first equation in (\ref{SEGB3N}), we find
\begin{align}
\label{SEGB11}
V(\phi(N)) =&\, \frac{3}{\kappa^2}H(N)^2 - \frac{1}{2}H(N)^2 \phi' (N)^2 - 3\e^N H(N)
\int^N \frac{dN_1}{\e^{N_1}} \left(\frac{2}{\kappa^2} H' (N_1) + H(N_1) \phi'(N_1)^2 \right)\, .
\end{align}
Eqs.~(\ref{SEGB10}) and (\ref{SEGB11}) and  tell that by using functions $h(N)$ and $ f(\phi)$, if $\xi(\phi)$ and $V(\phi)$ are given by
\begin{align}
\label{SEGB12}
V(\phi) =&\, \frac{3}{\kappa^2}h \left( f(\phi)\right)^2
 - \frac{h\left( f\left(\phi\right)\right)^2}{2 f'(\phi)^2} - 3h\left( f(\phi)\right) \e^{ f(\phi)}
\int^\phi d\phi_1  f'( \phi_1 ) \e^{- f(\phi_1)} \left(\frac{2}{\kappa^2}h'\left( f(\phi_1)\right)
+ \frac{h'\left( f\left(\phi_1\right)\right)}{ f'(\phi_1 )^2} \right)\, , \\
\label{SEGB13}
\xi(\phi) =&\, \frac{1}{8}\int^\phi d\phi_1
\frac{ f'(\phi_1) \e^{ f(\phi_1)} }{h\left( f\left(\phi_1\right)\right)^3}
\int^{\phi_1} d\phi_2  f'(\phi_2) \e^{- f(\phi_2)} \left(\frac{2}{\kappa^2}h'\left( f(\phi_2)\right)
+ \frac{h\left( f(\phi_2)\right)}{ f'(\phi_2)^2} \right)\, ,
\end{align}
a solution of the equations in (\ref{SEGB3}) is given by,
\begin{align}
\label{SEGB14}
\phi= f^{-1}(N)\quad \left(N= f(\phi)\right)\, ,\quad H = h(N) \, .
\end{align}
Therefore we obtain a general Einstein-Gauss-Bonnet model realizing the time-evolution of $H$
given by an arbitrary function $h(N)$ as in (\ref{SEGB14}).
We often solve the model with given potential, etc. but here, we have considered
the solution $H=h(N)$ first and we have constructed the model that realizes the given solution $H=h(N)$.
We should also note that the time-evolution is determined only by one function $h(N)$ but in the action (\ref{I8one}),
there appear two functions $V(\phi)$ and $\xi(\phi)$.
There is one additional functional degree of freedom in the model compared with the evolution.
We can separate the two functional degrees of freedom in the action into the function $h(N)$ relevant to the evolution
of $H$ and the function $f(\phi)$ irrelevant to the evolution.
Therefore if we change $f(\phi)$, the functional form of $\xi(\phi)$ changes but, of course, the functional form of $V(\phi)$ also changes.
The changes of $\xi(\phi)$ and $V(\phi)$ compensate with each other and the time evolution of the expansion of the Universe given by $H=h(N)$ does not change.

We should note again that the history of the expansion of the Universe
is determined only by the function $h(N)$ and does not depend on the choice of $f(\phi)$.
By using this indefiniteness of the choice of $f(\phi)$, we consider the possibility that $\xi$ goes
to a constant or vanishes in the late Universe.
The possibility can be satisfied if,
\begin{align}
\label{p2A1}
0 \sim \frac{2}{\kappa^2}H' (N) + H(N) {\phi'(N)}^2 \, ,
\end{align}
which can be solved as,
\begin{align}
\label{p2A2}
\phi (N) = f^{-1}(N) \sim \int dN \sqrt{ - \frac{2H'(N)}{\kappa^2 H(N)}}\, .
\end{align}
The above expression can be valid as long as $H'(N)<0$, which
corresponds to the case that the effective equation of state
parameter, which is defined by,
\begin{align}
\label{EC3_0}
w_\mathrm{eff} \equiv -1 - \frac{2 H'}{3 H} \, ,
\end{align}
is greater than $-1$.
Compared with the past Universe, in the late Universe, the Hubble
rate $H$ goes to a constant, and we expect that $H$ become
asymptotically constant in the future, that is, the spacetime
becomes an asymptotically de Sitter spacetime.
This feature indicates that if the scalar field $\phi$ goes to a constant in the late Universe, the condition (\ref{p2A2}) is satisfied.
This is natural because $\xi$ is a function of $\phi$, if $\phi$ goes to a constant, $\xi=\xi(\phi)$
becomes a constant as long as $\xi$ is not a singular function.

For example, we may assume,
\begin{align}
\label{xi1}
\xi = \xi_0 \left( 1 - \e^{-\xi_1 N} \right)\, .
\end{align}
Here $\xi_0$ and $\xi_1$ are constants and we assume $\xi_1$ is
positive. Then $\xi$ rapidly goes to a constant $\xi\to \xi_0$
when $N$ becomes large. And therefore, the Einstein-Gauss-Bonnet
gravity transits to the standard scalar-tensor theory. The cosmic
time of the transition can be adjusted by fine-tuning the
parameter $\xi_1$. For example, if we choose $1/\xi_1 \lesssim
60$, $\xi(\phi)$ becomes almost constant in the epoch of the
reheating. Eq.~(\ref{SEGB10}) indicates that,
\begin{align}
\label{SEGB10BB}
{\phi'}^2 = - \frac{2 H'}{\kappa^2 H} + 8 \xi_0
\xi_1\left\{ 2HH' - \left( \xi_1 + 1\right) H^2 \right\} \e^{-\xi_1 N}\, .
\end{align}
The second term decreases rapidly due to the factor $\e^{- \xi_1 N}$
and the first term vanishes, and therefore $\phi$ also goes to
a constant consistently if the spacetime becomes asymptotically a de Sitter spacetime.

We may estimate the propagating speed $c_\mathrm{GW}$ of the gravitational waves.
We now consider the plain wave
$h_{ij} \propto \mathrm{Re} \left( \e^{-i \omega t + i \bm{k}\cdot \bm{x}} \right)$.
Under the condition (\ref{gfc}) and (\ref{ce}), by using
(\ref{xis}), Eq.~(\ref{second}) gives the following dispersion
relation for high energy gravitational waves,
\begin{align}
\label{GWSFLRW1}
0 =  \frac{1}{2} \left( - \frac{1}{2\kappa^2} + 4 \left( \ddot \xi + 3 H \dot \xi\right) \right) \left( \omega^2 - \frac{k^2}{a^2} \right)
+ 2 \ddot \xi \omega^2 \, .
\end{align}
Here $k^2 = \bm{k}\cdot \bm{k}$.
Eq.~(\ref{GWSFLRW1}) tells that the
\begin{align}
\label{GWSFLRW2}
{c_\mathrm{GW}}^2 = \frac{1 - 8 \kappa^2 \left( \ddot \xi + 3 H \dot \xi\right)}{1 - 8\kappa^2 \left( 2 \ddot \xi
+ 3 H \dot \xi\right)} c^2 \sim \left( 1 + 8 \kappa^2 \ddot \xi \right) c^2 \, .
\end{align}
Now the speed of light $c$ is given by $c^2 = \frac{1}{a^2}$.
We also we assumed $\left| \kappa^2 \ddot \xi  \right| \ll 1$.
Eq.~(\ref{GWSFLRW2}) tells that if $\ddot\xi>0$ $\left(\ddot \xi<0\right)$,
the propagating speed of the gravitational wave is larger (smaller) than that of light.
The observation of GW170817 in (\ref{GWp9}) gives the following constraint,
\begin{align}
\label{GWp9FLRW}
\left| 8 \kappa^2 \ddot \xi \right| < 6 \times 10^{-15}\, .
\end{align}
Especially in the case of (\ref{xi1}), we obtain
\begin{align}
\label{GWp9FLRW}
\left| 8 \kappa^2 \xi_0 \left( \xi_1 H H' - {\xi_1}^2 H^2 \right) \e^{-\xi_1 N} \right| < 6 \times 10^{-15}\, .
\end{align}
In the present Universe, where the speed of the gravitational wave
was measured, we may assume $N=120\ - \ 140$.

We consider the following model in terms of $e$-foldings number
$N$, which satisfies the above conditions,
\begin{align}
\label{Ex2}
H=h(N) =H_0 \left( 1 + \alpha N^\beta \right)^\gamma \, .
\end{align}
We construct this model to describe the whole history of the
Universe, that is, inflation, matter-dominant epoch, and the
accelerating expansion of the present Universe.
When $N$ is small, we find $H \sim H_0 \left( 1 + \alpha \gamma N^\beta \right)$
and when $N$ is large, $H \sim H_1  \alpha^\gamma N^{\beta\gamma}$.
Therefore $H_1$ and $\alpha$ should be positive so that $H$ is positive.
We also require $\beta>0$ and $\gamma<0$ so that $H$ is a monotonically decreasing function.

The effective equation of state parameter is given by,
\begin{align}
\label{EC3}
w_\mathrm{eff} = - 1 - \frac{2 \alpha \beta \gamma N^{\beta - 1}}{3 \left( 1 + \alpha N^\beta \right) } \, ,
\end{align}
which goes to $-1$ when $N\to 0$ or $N\to \infty$,
We should note that $N\to 0$ corresponds to the early Universe and
$N\to \infty$ corresponds to the present or future Universe.
Therefore because the effective equation of state parameter
$w_\mathrm{eff}$ goes to $- 1$ when $N\to 0$ and when $N\to
\infty$, the model (\ref{Ex2}) describes both the inflation and
the accelerating expansion in the late Universe.

We now check if the model (\ref{Ex2}) also describes the matter-dominated area.
When $w_\mathrm{eff} = - \frac{1}{3}$, we find
\begin{align}
\label{EC4}
 - \alpha \beta \gamma N^{\beta - 1} = 1 + \alpha N^\beta \, .
\end{align}
Let the two solutions of (\ref{EC4}) $N=N_1$ and $N=N_2$ $\left(0<N_1<N_2\right)$.
Then the period where $N<N_1$ corresponds to the inflation in the early Universe and the period
where $N>N_2$ to the accelerating expansion in the present Universe.
During $N_1<N<N_2$, we expect $w_\mathrm{eff}$ goes to vanish at least if we include the contribution of the matter which is dust.

The parameters $\alpha$ and $\gamma$ are given in
terms of $\beta$, $N_1$, and $N_2$, as follows,
\begin{align}
\label{EC4B}
\alpha = \frac{ {N_2}^{\beta-1} - {N_1}^{\beta-1}}{\left(N_1 N_2\right)^{\beta-1} \left( N_2 - N_1 \right)} \, , \quad
\gamma= - \frac{{N_2}^\beta - {N_1}^\beta}{\beta \left( {N_2}^{\beta-1} - {N_1}^{\beta-1} \right)}\, .
\end{align}
We should note that $\alpha$ and $\gamma$ are positive as long as
$\beta>1$ as we required. We now estimate the parameters $\alpha$,
$\beta$, and $\gamma$ in order to obtain realistic models
compatible with the constraint on the gravitational wave speed.
Let the beginning of the inflation correspond to $N=0$. Then the
end of the inflation corresponds to $N=N_1=60-70$ and the
recombination (clear up of the Universe) to $N=120-140$. The
redshift of the recombination is $z=1100$. Because $1+z = 1/a$,
where $a$ is the scale factor, we obtain $N_0-N=\ln \left( 1+ z
\right)$, where $N_0$ is the redshift of the present Universe. We
note $\ln 1,100 \sim 7$. The redshift corresponding to the
beginning of the accelerating expansion of the late Universe is
approximately $0.4$ and $\ln 1.4 \sim 0.3$. Therefore $N_2\sim
2N_1$. In order to estimate $\alpha$ and $\beta$ in (\ref{EC4B}),
we assume $N_2 = 2 N_1 = \mathcal{O}\left( 10^2 \right)$. Then we
find,
\begin{align}
\label{EC4B2}
\alpha = \frac{1 - 2^{1-\beta}}{2^{-\beta} {N_2}^\beta} \, , \quad
\gamma = - \frac{\left( 1 - 2^{-\beta} \right) N_2}{\beta \left( 1 - 2^{1 - \beta} \right)}\, .
\end{align}
In the early Universe, where $N\to 0$, Eq.~(\ref{Ex2}) has the following form,
\begin{align}
\label{EC4B3}
H \sim H_0 \left( 1 + \alpha \gamma N^\beta \right) \, .
\end{align}
Therefore, $H_0$ corresponds to the scale of inflation and we now
choose $H_0 \sim 10^{14}\, \mathrm{GeV} = 10^{23}\, \mathrm{eV}$.
On the other hand, when $N$ is large $\left( N \sim 10^2 \right)$,
which corresponds to the period of the accelerating expansion of
the present Universe, we find,
\begin{align}
\label{EC4B4}
H=H_0 \alpha^ \gamma N^{\beta \gamma} \, ,
\end{align}
which requires $\alpha^\gamma N^{\beta \gamma} \sim 10^{-56}$ because $H\sim 10^{-33}\, \mathrm{eV}$.
Because $N\sim N_2$, by using (\ref{EC4B2}), we find,
\begin{align}
\label{EC4B5}
\left( 2^\beta - 2 \right)^{- \frac{\left( 1 - 2^{-\beta} \right) N_2}{\beta \left( 1 - 2^{1 - \beta} \right)}}
\sim 10^{-56} \, .
\end{align}
When $\beta \to 1$, we find $\left( 2^\beta - 2 \right)^{- \frac{\left( 1 - 2^{-\beta} \right) N_2}
{\beta \left( 1 - 2^{1 - \beta} \right)}} \to 0$.
On the other hand, when $\beta=2$, we find
$\left( 2^\beta - 2 \right)^{- \frac{\left( 1 - 2^{-\beta} \right) N_2}{\beta \left( 1 - 2^{1 - \beta} \right)}}
= 2^{- \frac{3}{4}N_2 } \sim 10^{-22}\gg 10^{-56}$.
Therefore, there is a solution $\beta$ for Eq.~(\ref{EC4B5}) when $1<\beta<2$.
Then Eq.~(\ref{EC4B2}) tells $\alpha \sim O\left( 10^{-\left(2-4\right)} \right)$ and
$\gamma \sim \mathcal{O}\left( 10^2 \right)$, and therefore these parameters are not too small or large.

\subsection{Gravitational Waves During the Inflationary Era}

We now consider the propagating speed of the gravitational wave during the inflationary era.
The expression of the propagating speed of the gravitational wave in (\ref{GWSFLRW2}) is valid and
at the beginning of the inflation $N\sim 0$, we find,
\begin{align}
\label{inf1}
8\kappa^2 \ddot\xi = 8\kappa^2 \xi_0 \xi_1 \left( \dot H - \xi H^2 \right) \e^{-\xi_1 N}
\sim - 8\kappa^2 {\xi_0}^2 \xi_1 {H_0}^2 \, .
\end{align}
Here we have assumed that Eqs.~(\ref{xi1}) and (\ref{Ex2}) hold true.
On the other hand, at the end of the inflation, we obtain,
\begin{align}
\label{GWSFLRW2end}
{c_\mathrm{GW}}^2 \sim \left( 1 - 8 \kappa^2 \xi_0 {H_0}^2 \left( 1 + \alpha {N_1}^\beta \right)^{2\gamma}
\left( \xi_1 + {\xi_1}^2 \right) \e^{-\xi_1 N_1} \right) c^2 \, .
\end{align}
Here we have used (\ref{Ex2}) and (\ref{EC4}) for $N=N_1$.

The gravitational wave generated in the epoch of inflation has not been detected.
Therefore there still be the possibility that the propagating speed of the gravitational wave might be significantly different
from the speed of light.
As a working hypothesis, we now assume that the speed of the gravitational wave is smaller by
$10\%$ than that of light during inflation.
Then Eq.~(\ref{GWSFLRW2}) tells,
\begin{align}
\label{inf2}
8\kappa^2 {\xi_0}^2 \xi_1 {H_0}^2 \sim \frac{1}{10}\, .
\end{align}
If we assume $\xi_1 = \mathcal{O}(1)$, the factor $\e^{-\xi_1 N} $ in (\ref{GWSFLRW2end}) is very small,
\begin{align}
\label{inf3}
\e^{-\xi_1 N} \sim \e^{-60} \fallingdotseq 8.8\times 10^{-27}\, .
\end{align}
Therefore, we expect that the difference between the speed of the
gravitational wave and that of light can be neglected after the
inflationary era, including the epoch of the reheating, a scenario
which we discuss in the next subsection.

\subsection{Reheating Scenario}\label{Sec8}

In this subsection, we estimate the temperature and the
propagating speed of the gravitational wave in the epoch of reheating.
We expect that the inflationary will end when $N=N_1$,
which is one of the solutions of Eq.~(\ref{EC4}).
So far, we have neglected the contributions from the matter fluids.
After $N=N_1$, if the scalar field $\phi$ couples with matter, it could affect
the reheating era and the evolution of $H$ deviates from that in
Eq.~(\ref{Ex2}). In order to investigate the behavior of the
scalar field dynamics, we expand the quantity around $N=N_1$ as
follows,
\begin{align}
\label{re1}
N=N_1 + \delta N\, .
\end{align}
Then by using (\ref{xi1}) and  (\ref{Ex2}), we find,
\begin{align}
\label{re2}
\xi' \sim &\, \xi_0 \xi_1 \e^{- \xi_1 N_1} \left( 1 - \xi_1 \delta N \right) \, , \quad
H \sim H_0 \left( 1 + \alpha {N_1}^\beta \right)^\gamma \left( 1 - \delta N \right)
\, , \nonumber \\
H' \sim&\, \alpha \beta \gamma H_0 \left( 1 + \alpha {N_1}^\beta \right)^{\gamma-1} {N_1}^{\beta -1} \left\{ 1
+ \left( - \frac{\gamma - 1}{\gamma} + \beta - 1 \right) \frac{\delta N}{N_1} \right\} \, .
\end{align}
Here we have used Eq.~(\ref{EC4}) with $N=N_1$ and the scalar
potential reads,
\begin{align}
\label{SEGB11re} V \sim &\, V_0 - \left\{ \frac{6}{\kappa^2}
\left( 1 + \frac{ \alpha \beta \gamma}{\left( 1 + \alpha
{N_1}^\beta \right)^\gamma } \right) + \frac{2{\phi_0}^2 \e^{-
\frac{2N_1}{N_0}}}{N_0} \right\} H_0^2 \left( 1 + \alpha
{N_1}^\beta \right)^{2\gamma} \delta N \, ,
\end{align}
where $V_0$ and $\xi_1$ are constants of integration, which can be
determined by using (\ref{SEGB3}), which gives when $N=N_1$,
\begin{align}
\label{SEGB3BBB}
0=&\, - \left( \frac{3}{\kappa^2} - \frac{{\phi_0}^2 \e^{- \frac{2N_1}{N_0}}}{2 N_0} \right) H_0^2 \left( 1 + \alpha {N_1}^\beta \right)^{2\gamma}
+ V_0 + 24 H_0^4 \left( 1 + \alpha {N_1}^\beta \right)^{4\gamma} \xi_0 \xi_1 \e^{- \xi_1 N_1} \, ,\nonumber \\
0=&\, \frac{1}{\kappa^2} \left( 1 - \frac{ \alpha \beta \gamma {N_1}^{\beta -1}}{1 + \alpha {N_1}^\beta} \right) H_0^2 \left( 1 + \alpha {N_1}^\beta \right)^{2\gamma} - V_0
 - H_0^4 \left( 1 + \alpha {N_1}^\beta \right)^{4\gamma} \left( 16 + \frac{24 \alpha \beta \gamma {N_1}^{\beta -1}}{1 + \alpha {N_1}^\beta} \right) \xi_0 \xi_1 \e^{- \xi_1 N_1} \, ,
\end{align}
that is,
\begin{align}
\label{SEGB3BBBBB}
\xi_0 \xi_1 \e^{- \xi_1 N_1} =&\, \left\{ \frac{1}{\kappa^2} \left( 2 + \frac{ \alpha \beta \gamma {N_1}^{\beta -1}}{1 + \alpha {N_1}^\beta} \right)
 - \frac{{\phi_0}^2 \e^{- \frac{2N_1}{N_0}}}{2 N_0} \right\} H_0^{-2} \left( 1 + \alpha {N_1}^\beta \right)^{- 2\gamma}
\left( 8 - \frac{24 \alpha \beta \gamma {N_1}^{\beta -1}}{1 + \alpha {N_1}^\beta} \right)^{-1} \, ,\nonumber \\
V_0 =&\, \frac{1}{\kappa^2} \left( 1 - \frac{ \alpha \beta \gamma {N_1}^{\beta -1}}{1 + \alpha {N_1}^\beta} \right) H_0^2
\left( 1 + \alpha {N_1}^\beta \right)^{2\gamma} - H_0^2 \left( 1 + \alpha {N_1}^\beta \right)^{2\gamma}  \nonumber \\
&\, \times \left( 16 + \frac{24 \alpha \beta \gamma {N_1}^{\beta -1}}{1 + \alpha {N_1}^\beta} \right) \left\{ \frac{1}{\kappa^2}
\left( 2 + \frac{ \alpha \beta \gamma {N_1}^{\beta -1}}{1 + \alpha {N_1}^\beta} \right)
 - \frac{{\phi_0}^2 \e^{- \frac{2N_1}{N_0}}}{2 N_0} \right\} \left( 8 - \frac{24 \alpha \beta \gamma {N_1}^{\beta -1}}{1 + \alpha {N_1}^\beta} \right)^{-1} \, .
\end{align}
We now estimate the reheating temperature $T_\mathrm{re}$. The
effective energy density at the end of the inflationary era is
given by,
\begin{align}
\label{reT1}
\rho_\mathrm{eff} = \frac{3}{\kappa^2} H\left( N_1 \right)^2
= \frac{3}{\kappa^2} {H_0}^2 \left( 1 + \alpha {N_1}^\beta \right)^{2\gamma}\, .
\end{align}
We now assume that all the energy density is transformed into
radiation. The Stefan-Boltzmann law indicates that,
\begin{align}
\label{reT2}
\frac{3}{\kappa^2} {H_0}^2 \left( 1 + \alpha {N_1}^\beta \right)^{2\gamma}
= \left( \frac{\pi^2 g_\mathrm{re}}{30} \right) {T_\mathrm{re}}^4 \, .
\end{align}
Here $g_\mathrm{re}$ denotes the number of the massless degrees of
freedom when the reheating era occurred. Then we obtain,
\begin{align}
\label{reT3}
T_\mathrm{re} = \frac{3}{\kappa^2} \sqrt{H_0} \left( 1 + \alpha {N_1}^\beta \right)^\frac{\gamma}{2}
\left( \frac{30}{\pi^2 g_\mathrm{re}} \right)^\frac{1}{4} \, .
\end{align}
In the epoch of the reheating, by using (\ref{GWSFLRW2}), we find the propagating speed of the gravitational wave as follows,
\begin{align}
\label{GWSFLRW2rehe} {c_\mathrm{GW}}^2 \sim \left( 1 + 8 \kappa^2
\xi_0 {H_0}^2  \left( 1 + \alpha {N_\mathrm{rh}}^\beta
\right)^{2\gamma-1} \left( \xi_1 \alpha \beta \gamma
{N_\mathrm{rh}}^{\beta - 1} - {\xi_1}^2 \left( 1 + \alpha
{N_\mathrm{rh}}^\beta \right) \right) \e^{-\xi_1 N_\mathrm{rh}}
\right) c^2 \, ,
\end{align}
where $N_\mathrm{rh}$ is the $e$-folding number corresponding to
the reheating. So from the above relation, we have a concrete idea
on the behavior of the gravitational wave speed during the
reheating era, which is non-trivial, as expected, and somewhat
model dependent.


\section{Propagation of Gravitational Waves in Spherically Symmetric Spacetime}\label{SecIIInew}

In this section, we consider the propagation of the gravitational
waves in a spherically symmetric spacetime background.
First, we consider the spherically symmetric and also time-dependent spacetime.
The spacetime includes both the static spherically
symmetric spacetime and the FLRW spacetime as special cases.
We show that the condition (\ref{condition}) cannot be satisfied in
the non-trivial but general spherically symmetric spacetime.
After that, we estimate the deviation of the propagating speed of the
gravitational wave from the speed of light inside a stellar object.

\subsection{Spherically Symmetric Time-dependent Spacetime}\label{Sec4}

In this subsection, we show that the condition (\ref{condition})
cannot be satisfied in the non-trivial but general spherically symmetric spacetime.
The most general form of the spherically symmetric and time-dependent spacetime is given by,
\begin{equation}
\label{G1}
ds^2 = - \mathcal{A}(\tau,\rho) d\tau^2 + 2 \mathcal{B} (\tau,\rho) d\tau d\rho
+ \mathcal{C}(\tau,\rho) d\rho^2 + \mathcal{D} (\tau,\rho)
\left( d\theta^2 + \sin^2\theta d\phi^2 \right) \, .
\end{equation}
We should note that the spatially flat FLRW Universe is a special
class of the above spacetime. We define the radial coordinate $r$
by,
\begin{equation}
\label{G2}
r^2 \equiv \mathcal{D} (\tau,\rho) \, ,
\end{equation}
by assuming $\mathcal{D} (\tau,\rho)$ is positive.
In principle, Eq.~(\ref{G2}) can be solved with respect to $\rho$ as $\rho=\rho(\tau, r)$.
Then the metric in (\ref{G1}) can be rewritten as,
\begin{align}
\label{G3}
ds^2 =& \left\{ - \mathcal{A}\left(\tau,\rho\left(\tau,r\right) \right)
+ 2 \mathcal{B} \left(\tau,\rho\left(\tau,r\right) \right) \frac{\partial\rho}{\partial\tau}
\right\} d\tau^2
+ 2 \mathcal{B} \left(\tau,\rho\left(\tau,r\right) \right) \frac{\partial \rho}{\partial r}
d\tau dr \nonumber \\
& + \mathcal{C}\left(\tau,\rho\left(\tau,r\right) \right)
\left( \frac{\partial \rho}{\partial r} \right)^2 dr^2
+ r^2 \left( d\theta^2 + \sin^2\theta d\phi^2 \right) \, .
\end{align}
Furthermore, we introduce a new time coordinate $t$ as $\tau=\tau(t,r)$.
Then the metric in (\ref{G3}) can be further rewritten as,
\begin{align}
\label{G4}
ds^2 =& \left\{ - \mathcal{A}\left(\tau\left(t,r\right),
\rho\left(\tau\left(t,r\right),r\right) \right)
+ 2 \mathcal{B} \left(\tau\left(t,r\right),\rho\left(\tau\left(t,r\right),r\right)
\right) \frac{\partial\rho\left(\tau\left(t,r\right),r\right)}{\partial\tau}
\right\} \left( \frac{\partial \tau\left(t,r\right)}{\partial t} \right)^2 dt^2 \nonumber \\
& + 2 \left[ \mathcal{B} \left(\tau\left(t,r\right),\rho\left(\tau\left(t,r\right),r\right) \right)
\frac{\partial \rho\left(\tau\left(t,r\right),r\right)}{\partial r}
\frac{\partial \tau\left(t,r\right)}{\partial t} \right. \nonumber \\
& \left. + \left\{ - \mathcal{A}\left(\tau\left(t,r\right),
\rho\left(\tau\left(t,r\right),r\right) \right)
+ 2 \mathcal{B} \left(\tau\left(t,r\right),\rho\left(\tau\left(t,r\right),r\right)
\right) \frac{\partial\rho\left(\tau\left(t,r\right),r\right)}{\partial\tau}
\right\}
\frac{\partial \tau\left(t,r\right)}{\partial t}
\frac{\partial \tau\left(t,r\right)}{\partial r} \right] dt dr \nonumber \\
& + \left[ \mathcal{C}\left(\tau,\rho\left(\tau,r\right) \right)
\left( \frac{\partial \rho}{\partial r} \right)^2
+ \mathcal{B} \left(\tau\left(t,r\right),\rho\left(\tau\left(t,r\right),r\right) \right)
\frac{\partial \rho\left(\tau\left(t,r\right),r\right)}{\partial r}
\frac{\partial \tau\left(t,r\right)}{\partial r} \right. \nonumber \\
& \left. \left\{ - \mathcal{A}\left(\tau\left(t,r\right),
\rho\left(\tau\left(t,r\right),r\right) \right)
+ 2 \mathcal{B} \left(\tau\left(t,r\right),\rho\left(\tau\left(t,r\right),r\right)
\right) \frac{\partial\rho\left(\tau\left(t,r\right),r\right)}{\partial\tau}
\right\}
\left( \frac{\partial \tau\left(t,r\right)}{\partial r} \right)^2
\right] dr^2 \nonumber \\
& + r^2 \left( d\theta^2 + \sin^2\theta d\phi^2 \right) \, .
\end{align}
We can choose the time-coordinate $t$ so that,
\begin{align}
\label{G5}
0=& \mathcal{B} \left(\tau\left(t,r\right),\rho\left(\tau\left(t,r\right),r\right) \right)
\frac{\partial \rho\left(\tau\left(t,r\right),r\right)}{\partial r}
\frac{\partial \tau\left(t,r\right)}{\partial t} \nonumber \\
& + \left\{ - \mathcal{A}\left(\tau\left(t,r\right),
\rho\left(\tau\left(t,r\right),r\right) \right)
+ 2 \mathcal{B} \left(\tau\left(t,r\right),\rho\left(\tau\left(t,r\right),r\right)
\right) \frac{\partial\rho\left(\tau\left(t,r\right),r\right)}{\partial\tau}
\right\}
\frac{\partial \tau\left(t,r\right)}{\partial t}
\frac{\partial \tau\left(t,r\right)}{\partial r} \, .
\end{align}
Then finally, the metric has the following form,
\begin{align}
\label{GBiv}
ds^2 =&\, - \e^{2\nu (r,t)} dt^2 + \e^{2\lambda (r,t)} dr^2 + r^2 \left( d\theta^2 + \sin^2\theta d\phi^2 \right)\, , \nonumber \\
 - \e^{2\nu (r,t)} \equiv &\, \left\{ - \mathcal{A}\left(\tau\left(t,r\right), \rho\left(\tau\left(t,r\right),r\right) \right)
+ 2 \mathcal{B} \left(\tau\left(t,r\right),\rho\left(\tau\left(t,r\right),r\right) \right) \frac{\partial\rho\left(\tau\left(t,r\right),r\right)}{\partial\tau}
\right\} \left( \frac{\partial \tau\left(t,r\right)}{\partial t} \right)^2 \, , \nonumber \\
\e^{2\lambda (r,t)} \equiv\, & \mathcal{C}\left(\tau,\rho\left(\tau,r\right) \right) \left( \frac{\partial \rho}{\partial r} \right)^2
+ \mathcal{B} \left(\tau\left(t,r\right),\rho\left(\tau\left(t,r\right),r\right) \right)
\frac{\partial \rho\left(\tau\left(t,r\right),r\right)}{\partial r} \frac{\partial \tau\left(t,r\right)}{\partial r} \nonumber \\
&\, \left\{ - \mathcal{A}\left(\tau\left(t,r\right), \rho\left(\tau\left(t,r\right),r\right) \right)
+ 2 \mathcal{B} \left(\tau\left(t,r\right),\rho\left(\tau\left(t,r\right),r\right)
\right) \frac{\partial\rho\left(\tau\left(t,r\right),r\right)}{\partial\tau}
\right\} \left( \frac{\partial \tau\left(t,r\right)}{\partial r} \right)^2 \, .
\end{align}
We define the metric $\tilde g_{ij}$ of the unit sphere by
$\sum_{i,j=1,2} \tilde g_{ij} dx^i dx^j = d\theta^2 + \sin^2\theta d\phi^2$.
For the metric (\ref{GBiv}), the non-vanishing connections are the following,
\begin{align}
\label{GBv0}
&\Gamma^t_{tt}=\dot\nu \, , \quad \Gamma^r_{tt} = \e^{-2\lambda + 2\nu}\nu' \, ,
\quad \Gamma^t_{tr}=\Gamma^t_{rt}=\nu'\, , \quad
\Gamma^t_{rr} = \e^{2\lambda - 2\nu}\dot\lambda \, , \quad
\Gamma^r_{tr} = \Gamma^r_{rt} = \dot\lambda \, , \quad
\Gamma^r_{rr}=\lambda'\, ,\nonumber \\
&\Gamma^i_{jk} = \tilde \Gamma^i_{jk}\, ,\quad \Gamma^r_{ij}=-\e^{-2\lambda}r\tilde g_{ij} \, ,
\quad \Gamma^i_{rj}=\Gamma^i_{jr}=\frac{1}{r}\delta^i_{\ j}\, .
\end{align}
Here $\tilde \Gamma^i_{jk}$ is the connection given by $\tilde g_{ij}$.
Since,
\begin{equation}
\label{Riemann}
R^\lambda_{\ \mu\rho\nu}=
 -\Gamma^\lambda_{\mu\rho,\nu}
+ \Gamma^\lambda_{\mu\nu,\rho}
 - \Gamma^\eta_{\mu\rho}\Gamma^\lambda_{\nu\eta}
+ \Gamma^\eta_{\mu\nu}\Gamma^\lambda_{\rho\eta} \, ,
\end{equation}
we find that,
\begin{align}
\label{curvatures}
R_{rtrt} =&\, - \e^{2\lambda} \left\{ \ddot\lambda
+ \left( \dot\lambda - \dot\nu \right) \dot\lambda \right\}
+ \e^{2\nu}\left\{\nu'' + \left(\nu' - \lambda'\right)\nu' \right\} \, ,\quad
R_{titj} = r\nu'\e^{2(\nu - \lambda)} \tilde g_{ij} \, ,\nonumber \\
R_{rirj} =&\, \lambda' r \tilde g_{ij} \, ,\quad
R_{tirj}= \dot\lambda r \tilde g_{ij} \, , \quad
R_{ijkl} = \left( 1 - \e^{-2\lambda}\right) r^2
\left(\tilde g_{ik} \tilde g_{jl} - \tilde g_{il} \tilde g_{jk} \right)\, ,\nonumber \\
R_{tt}=&\, - \left\{ \ddot\lambda
+ \left( \dot\lambda - \dot\nu \right) \dot\lambda \right\}
+ \e^{2\left(\nu - \lambda\right)} \left\{
\nu'' + \left(\nu' - \lambda'\right)\nu' + \frac{2\nu'}{r}\right\} \, ,\nonumber \\
R_{rr} =&\, \e^{-2\left( \nu - \lambda \right)} \left\{ \ddot\lambda
+ \left( \dot\lambda - \dot\nu \right) \dot\lambda \right\}
 - \left\{ \nu'' + \left(\nu' - \lambda'\right)\nu' \right\}
+ \frac{2 \lambda'}{r} \, ,\nonumber \\
R_{tr} =&\, \frac{2\dot\lambda}{r} \, , \quad
R_{ij} = \left[ 1 + \left\{ - 1 - r \left(\nu' - \lambda' \right)\right\}\e^{-2\lambda}\right]
\tilde g_{ij}\ , \nonumber \\
R=&\, 2 \e^{-2 \nu} \left\{ \ddot\lambda
+ \left( \dot\lambda - \dot\nu \right) \dot\lambda \right\}
+ \e^{-2\lambda}\left[ - 2\nu'' - 2\left(\nu'  - \lambda'\right)\nu' - \frac{4\left(\nu' - \lambda'\right)}{r} + \frac{2\e^{2\lambda} - 2}{r^2} \right] \, .
\end{align}
By assuming that $\xi$ only depends on $r$ and $t$ because we are
considering spherically symmetric spacetime, we find,
\begin{align}
\label{nablaxi}
\nabla_t \nabla_t \xi =&\, {\partial_t}^2 \xi - \dot\nu \partial_t \xi - \e^{-2\lambda + 2\nu}\nu' \partial_r \xi \, , \quad
\nabla_r \nabla_r \xi = {\partial_r}^2 \xi - \e^{2\lambda - 2\nu}\dot\lambda \partial_t \xi - \lambda' \partial_r \xi \, , \nonumber \\
\nabla_i \nabla_j \xi =&\, \e^{-2\lambda}r\tilde g_{ij} \partial_r \xi \, , \quad
\nabla_r \nabla_t \xi = \nabla_t \nabla_r \xi = \partial_r \partial_t \xi - \nu' \partial_t \xi - \dot\lambda \partial_r \xi \, , \nonumber \\
\nabla_t \nabla_i \xi = &\, \nabla_i \nabla_t \xi = \nabla_r \nabla_i \xi = \nabla_i \nabla_r \xi = 0 \, , \nonumber \\
\nabla^2 \xi =&\, - \e^{-2\nu} \left( {\partial_t}^2 \xi - \dot\nu \partial_t \xi - \e^{-2\lambda + 2\nu}\nu' \partial_r \xi \right)
+ \e^{-2\lambda}\left(  {\partial_r}^2 \xi - \e^{2\lambda - 2\nu}\dot\lambda \partial_t \xi - \lambda' \partial_r \xi \right)
+ \frac{2\e^{-2\lambda}}{r} \partial_r \xi \, .
\end{align}
Then the condition (\ref{condition}) gives,
\begin{align}
\label{conditions}
0=&\, 3 \e^{-2\nu} \left( {\partial_t}^2 \xi - \dot\nu \partial_t \xi - \e^{-2\lambda + 2\nu}\nu' \partial_r \xi \right)
+ \e^{-2\lambda}\left(  {\partial_r}^2 \xi - \e^{2\lambda - 2\nu}\dot\lambda \partial_t \xi - \lambda' \partial_r \xi \right)
+ \frac{2\e^{-2\lambda}}{r} \partial_r \xi \, , \nonumber \\
0=&\, - \e^{-2\nu} \left( {\partial_t}^2 \xi - \dot\nu \partial_t \xi - \e^{-2\lambda + 2\nu}\nu' \partial_r \xi \right)
 - 3 \e^{-2\lambda}\left(  {\partial_r}^2 \xi - \e^{2\lambda - 2\nu}\dot\lambda \partial_t \xi - \lambda' \partial_r \xi \right)
+ \frac{2\e^{-2\lambda}}{r} \partial_r \xi \, , \nonumber \\
0=&\, \partial_r \partial_t \xi - \nu' \partial_t \xi - \dot\lambda \partial_r \xi \, .
\end{align}
By combining the first and second equations in (\ref{conditions}),
we obtain,
\begin{align}
\label{conditionA1}
0=&\, \e^{\nu + \lambda} \left\{ \e^{-2\nu} \partial_t \left( \e^{-\nu - \lambda} \partial_t \xi \right)
+ \e^{- 2\lambda} \partial_r \left( \e^{-\nu - \lambda} \partial_r \xi \right) \right\}
\, . \\
\label{conditionA2}
0=&\, - \left(  {\partial_r}^2 \xi - \e^{2\lambda - 2\nu}\dot\lambda \partial_t \xi - \lambda' \partial_r \xi \right)
+ \frac{1}{r} \partial_r \xi \, , \\
\label{conditionA3}
0=&\, \left( {\partial_t}^2 \xi - \dot\nu \partial_t \xi - \e^{-2\lambda + 2\nu}\nu' \partial_r \xi \right)
+ \frac{\e^{-2\lambda+2\nu}}{r} \partial_r \xi \, .
\end{align}
For simplicity, we consider the case that the spacetime is static,
that is, $\nu$ and $\lambda$ do not depend on time coordinate $t$.
Then the equations in (\ref{conditions}) reduce to,
\begin{align}
\label{conditions2}
0=&\, 3 \e^{-2\nu} \left( {\partial_t}^2 \xi - \e^{-2\lambda + 2\nu}\nu' \partial_r \xi \right)
+ \e^{-2\lambda}\left(  {\partial_r}^2 \xi - \lambda' \partial_r \xi \right)
+ \frac{2\e^{-2\lambda}}{r} \partial_r \xi \, , \nonumber \\
0=&\, - \e^{-2\nu} \left( {\partial_t}^2 \xi - \e^{-2\lambda + 2\nu}\nu' \partial_r \xi \right)
 - 3 \e^{-2\lambda}\left(  {\partial_r}^2 \xi - \lambda' \partial_r \xi \right)
+ \frac{2\e^{-2\lambda}}{r} \partial_r \xi \, , \nonumber \\
0=&\, \partial_r \partial_t \xi - \nu' \partial_t \xi \, ,
\end{align}
and Eqs.~(\ref{conditionA2}) and (\ref{conditionA3}) reduce to,
\begin{align}
\label{conditionA2s}
0=&\, - \left(  {\partial_r}^2 \xi - \lambda' \partial_r \xi \right) + \frac{1}{r} \partial_r \xi \, , \\
\label{conditionA3s}
0=&\, {\partial_t}^2 \xi - \e^{-2\lambda + 2\nu}\nu' \partial_r \xi + \frac{\e^{-2\lambda+2\nu}}{r} \partial_r \xi \, .
\end{align}
The general solution of the last equation in (\ref{conditions2})
is given by,
\begin{align}
\label{sol1}
\xi (t, r) = \xi_{(t)}(t) \e^{\nu(r)} + \xi_{(r)}(r)\, ,
\end{align}
Here $\xi_{(t)}$ and $\xi_{(r)}$ are arbitrary functions of $t$ and $r$, respectively.
By substituting (\ref{sol1}) into (\ref{conditionA2s}), we obtain,
\begin{align}
\label{conditionA2s2}
0= - \left( \nu'' + {\nu'}^2 \right) \xi_{(t)} \e^\nu - \xi_{(r)}'' + \left(\lambda' + \frac{1}{r} \right) \left( \nu' \xi_{(t)} \e^\nu + \xi_{(r)}' \right) \, ,
\end{align}
which gives,
\begin{align}
\label{con1}
0 = - \nu'' +{\nu'}^2 + \left(\lambda' + \frac{1}{r} \right) \nu' \, , \quad
0 = - \xi_{(r)}'' + \left(\lambda' + \frac{1}{r} \right) \xi_{(r)}' \, .
\end{align}
The first equation in (\ref{con1}) gives a non-trivial relation for the spacetime geometry,
\begin{align}
\label{sol2}
0 = - \ln \frac{\nu'}{\nu'_0 r} + \nu + \lambda \, .
\end{align}
Here $\nu'_0$ is an integration constant.
On the other hand, the second equation in (\ref{con1}) can be solved as follows,
\begin{align}
\label{sol3}
\xi_{(r)}' = \xi_0 r \e^\lambda \, .
\end{align}
Here $\xi_0$ is an integration constant.
By substituting Eq.~(\ref{sol1}) and (\ref{sol3}) into (\ref{conditionA3s}) and by
using Eq.~(\ref{sol2}), we obtain,
\begin{align}
\label{con2}
0 = {\ddot\xi}_{(t)} \e^\nu - \left( \nu' - \frac{1}{r} \right) \frac{{\nu'_0}^2}{{\nu'}^2} \e^{4\nu} \left( \nu' \xi_{(t)} \e^\nu + \xi_{(r)}' \right) \, ,
\end{align}
which gives
\begin{align}
\label{con3}
0= {\ddot\xi}_{(t)} \, , \quad \nu'=\frac{1}{r}\, .
\end{align}
which yields,
\begin{align}
\label{cons4} \xi_{(t)}=\xi_1 \, , \quad 0= \nu' \xi_1 \e^\nu+
\xi_{(r)}'\, ,
\end{align}
where $\xi_1$ is a constant.
The second equation in (\ref{con3}) gives,
\begin{align}
\label{I2BH4}
\nu = \ln \frac{r}{r_0}\, ,
\end{align}
where $r_0$ is a constant.
On the other hand, when Eq.~(\ref{cons4}) is satisfied, Eq.~(\ref{sol1}) indicates that
$\xi$ does not depend on the time coordinate $t$.
Then Eq.~(\ref{conditionA3s}) yields,
\begin{align}
\label{cons5}
\nu'=\frac{1}{r}\, ,
\end{align}
which gives (\ref{I2BH4}), again.
Eq.~(\ref{I2BH4}) indicates that there is no horizon and therefore there is no solution for the
black hole when the speed of the propagating speed exactly
coincides with that of the light, even if we include two scalar
fields $\phi$ and $\chi$ in addition to matter.
Eq.~(\ref{cons5}) also prohibits more general but non-trivial spherically symmetric
spacetime, including the stellar configuration and wormholes.

In the next  subsection, we try to solve the problem of the
propagating speed of gravitational waves, by considering a model
of Einstein-Gauss-Bonnet gravity coupled with two scalar fields.
As we will show we reobtain the condition (\ref{condition}) again,
and therefore the propagating speed of the gravitational wave does
not coincide with that of light.


\section{Two-Scalar Einstein-Gauss-Bonnet Gravity}\label{Sec2}

Since the propagating speed of the gravitational waves cannot be
equal to that of light in the non-trivial spherically symmetric
background as we have shown in \cite{Nojiri:2023jtf}, we consider
the model including two scalar fields $\phi$ and $\chi$ to
investigate if the problem could be solved or not.

The action with two scalar fields is given by,
\begin{align}
\label{I8} S_{\phi\chi} = \int d^4 x \sqrt{-g} & \left\{
\frac{R}{2\kappa^2}
 - \frac{1}{2} A (\phi,\chi) \partial_\mu \phi \partial^\mu \phi
 - B (\phi,\chi) \partial_\mu \phi \partial^\mu \chi \right. \nonumber \\
& \left. \qquad - \frac{1}{2} C (\phi,\chi) \partial_\mu \chi
\partial^\mu \chi
 - V (\phi,\chi) - \xi(\phi, \chi) \mathcal{G} + \mathcal{L}_\mathrm{matter} \right\}\, .
\end{align}
Here $V(\phi,\chi)$ is the potential for $\phi$ and $\chi$ and $\xi(\phi,\chi)$ is also a function of $\phi$ and $\chi$.
By varying the action (\ref{I8}) with respect to the metric
$g_{\mu\nu}$, we obtain,
\begin{align}
\label{gb4bD4}
0= &\, \frac{1}{2\kappa^2}\left(- R_{\mu\nu} + \frac{1}{2} g_{\mu\nu} R\right) \nonumber \\
&\, + \frac{1}{2} g_{\mu\nu} \left\{
 - \frac{1}{2} A (\phi,\chi) \partial_\rho \phi \partial^\rho \phi
 - B (\phi,\chi) \partial_\rho \phi \partial^\rho \chi
 - \frac{1}{2} C (\phi,\chi) \partial_\rho \chi \partial^\rho \chi - V (\phi,\chi)\right\} \nonumber \\
&\, + \frac{1}{2} \left\{ A (\phi,\chi) \partial_\mu \phi
\partial_\nu \phi + B (\phi,\chi) \left( \partial_\mu \phi
\partial_\nu \chi + \partial_\nu \phi \partial_\mu \chi \right)
+ C (\phi,\chi) \partial_\mu \chi \partial_\nu \chi \right\} \nonumber \\
&\, - 2 \left( \nabla_\mu \nabla_\nu \xi(\phi,\chi)\right)R + 2
g_{\mu\nu} \left( \nabla^2 \xi(\phi,\chi)\right)R + 4 \left(
\nabla_\rho \nabla_\mu \xi(\phi,\chi)\right)R_\nu^{\ \rho}
+ 4 \left( \nabla_\rho \nabla_\nu \xi(\phi,\chi)\right)R_\mu^{\ \rho} \nonumber \\
&\, - 4 \left( \nabla^2 \xi(\phi,\chi) \right)R_{\mu\nu}
 - 4g_{\mu\nu} \left( \nabla_\rho \nabla_\sigma \xi(\phi,\chi) \right) R^{\rho\sigma}
+ 4 \left(\nabla^\rho \nabla^\sigma \xi(\phi,\chi) \right)
R_{\mu\rho\nu\sigma} + \frac{1}{2} T_{\mathrm{matter}\, \mu\nu} \, ,
\end{align}
and the field equation for the scalar field is obtained by varying
the action with respect to $\phi$ and $\chi$, and it is equal to,
\begin{align}
\label{I10} 0 =& \frac{1}{2} A_\phi \partial_\mu \phi \partial^\mu
\phi + A \nabla^\mu \partial_\mu \phi + A_\chi \partial_\mu \phi
\partial^\mu \chi + \left( B_\chi - \frac{1}{2} C_\phi
\right)\partial_\mu \chi \partial^\mu \chi
+ B \nabla^\mu \partial_\mu \chi - V_\phi - \xi_\phi \mathcal{G} \, ,\nonumber \\
0 =& \left( - \frac{1}{2} A_\chi + B_\phi \right) \partial_\mu
\phi \partial^\mu \phi + B \nabla^\mu \partial_\mu \phi
+ \frac{1}{2} C_\chi \partial_\mu \chi \partial^\mu \chi
+ C \nabla^\mu \partial_\mu \chi + C_\phi \partial_\mu \phi
\partial^\mu \chi
 - V_\chi - \xi_\chi \mathcal{G} \, .
\end{align}
Here $A_\phi=\partial A(\phi,\chi)/\partial \phi$, and similar
notation is used in other functions.
Also in (\ref{gb4bD4}), $T_{\mathrm{matter}\, \mu\nu}$ is the energy-momentum tensor of
perfect matter fluids.
We should note that the field equations in (\ref{I10}) are nothing but the Bianchi identities.
We again consider the equation that describes the gravitational waves and
we obtain the condition that the propagating speed of the
gravitational wave is equal to that of light.

By considering the variation of the metric in (\ref{variation1}),
we obtain the equation describing the propagation of the
gravitational waves as follows,
\begin{align}
\label{gb4bD4Btwo}
0=&\, \left[ \frac{1}{4\kappa^2} R + \frac{1}{2}\left\{
 - \frac{1}{2} A \partial_\rho \phi \partial^\rho \phi
 - B \partial_\rho \phi \partial^\rho \chi
 - \frac{1}{2} C \partial_\rho \chi \partial^\rho \chi - V \right\}
 - 4 \left( \nabla_{\rho} \nabla_\sigma \xi \right) R^{\rho\sigma} \right] h_{\mu\nu} \nonumber \\
&\, + \bigg[ \frac{1}{4} g_{\mu\nu} \left\{
 - A \partial^\tau \phi \partial^\eta \phi
 - B \left(\partial^\tau \phi \partial^\eta \chi + \partial^\eta \phi \partial^\tau \chi \right)
 - C \partial^\tau \chi \partial^\eta \chi \right\} \nonumber \\
&\, - 2 g_{\mu\nu} \left( \nabla^\tau \nabla^\eta \xi\right)R
 - 4 \left( \nabla^\tau \nabla_\mu \xi\right)R_\nu^{\ \eta} - 4 \left( \nabla^\tau \nabla_\nu \xi\right)R_\mu^{\ \eta}
+ 4 \left( \nabla^\tau \nabla^\eta \xi \right)R_{\mu\nu} \nonumber \\
&\, + 4g_{\mu\nu} \left( \nabla^\tau \nabla_\sigma \xi \right)
R^{\eta\sigma} + 4g_{\mu\nu} \left( \nabla_\rho \nabla^\tau \xi
\right) R^{\rho\eta}
 - 4 \left(\nabla^\tau \nabla^\sigma \xi \right) R_{\mu\ \, \nu\sigma}^{\ \, \eta}
 - 4 \left(\nabla^\rho \nabla^\tau \xi \right) R_{\mu\rho\nu}^{\ \ \ \ \eta}
\bigg] h_{\tau\eta} \nonumber \\
&\, + \frac{1}{2}\left\{ 2 \delta_\mu^{\ \eta} \delta_\nu^{\
\zeta} \left( \nabla_\kappa \xi \right)R
 - 2 g_{\mu\nu} g^{\eta\zeta} \left( \nabla_\kappa \xi \right)R
 - 4 \delta_\rho^{\ \eta} \delta_\mu^{\ \zeta} \left( \nabla_\kappa \xi \right)R_\nu^{\ \rho}
 - 4 \delta_\rho^{\ \eta} \delta_\nu^{\ \zeta} \left( \nabla_\kappa \xi \right)R_\mu^{\ \rho} \right. \nonumber \\
&\, \left. + 4 g^{\eta\zeta} \left( \nabla_\kappa \xi \right)
R_{\mu\nu} + 4g_{\mu\nu} \delta_\rho^{\ \eta} \delta_\sigma^{\
\zeta} \left( \nabla_\kappa \xi \right) R^{\rho\sigma}
 - 4 g^{\rho\eta} g^{\sigma\zeta} \left( \nabla_\kappa \xi \right) R_{\mu\rho\nu\sigma}
\right\} g^{\kappa\lambda}\left( \nabla_\eta h_{\zeta\lambda} + \nabla_\zeta h_{\eta\lambda} - \nabla_\lambda h_{\eta\zeta} \right) \nonumber \\
&\, + \left\{ \frac{1}{4\kappa^2} g_{\mu\nu} - 2 \left( \nabla_\mu
\nabla_\nu \xi\right) + 2 g_{\mu\nu} \left( \nabla^2\xi\right)
\right\}
\left\{ -h_{\mu\nu} R^{\mu\nu} + \nabla^\mu \nabla^\nu h_{\mu\nu} - \nabla^2 \left(g^{\mu\nu}h_{\mu\nu}\right) \right\} \nonumber \\
&\, + \frac{1}{2}\left\{ \left( - \frac{1}{2\kappa^2} - 4 \nabla^2
\xi \right) \delta^\tau_{\ \mu} \delta^\eta_{\ \nu} + 4 \left(
\nabla_\rho \nabla_\mu \xi\right) \delta^\eta_{\ \nu} g^{\rho\tau}
+ 4 \left( \nabla_\rho \nabla_\nu \xi\right) \delta^\tau_{\ \mu}
g^{\rho\eta}
 - 4g_{\mu\nu} \nabla^\tau \nabla^\eta \xi \right\} \nonumber \\
&\, \qquad \times \left\{\nabla_\tau\nabla^\phi h_{\eta\phi} +
\nabla_\eta \nabla^\phi h_{\tau\phi} - \nabla^2 h_{\tau\eta}
 - \nabla_\tau \nabla_\eta \left(g^{\phi\lambda}h_{\phi\lambda}\right)
 - 2R^{\lambda\ \phi}_{\ \eta\ \tau}h_{\lambda\phi}
+ R^\phi_{\ \tau}h_{\phi\eta} + R^\phi_{\ \tau}h_{\phi\eta} \right\} \nonumber \\
&\, + 2 \left(\nabla^\rho \nabla^\sigma \xi \right) \left\{
\nabla_\nu \nabla_\rho h_{\sigma\mu}
 - \nabla_\nu \nabla_\mu h_{\sigma\rho}
 - \nabla_\sigma \nabla_\rho h_{\nu\mu}
 + \nabla_\sigma \nabla_\mu h_{\nu\rho}
+ h_{\mu\phi} R^\phi_{\ \rho\nu\sigma}
 - h_{\rho\phi} R^\phi_{\ \mu\nu\sigma} \right\} \nonumber \\
&\, + \frac{1}{2}\frac{\partial T_{\mathrm{matter}\,
\mu\nu}}{\partial g_{\tau\eta}}h_{\tau\eta} \, .
\end{align}
Here we have assumed that the matter fluids minimally couple with gravity, once more.
By choosing the conditions in (\ref{gfc}) and
(\ref{ce}), we can reduce Eq.~(\ref{gb4bD4Btwo}) as follows,
\begin{align}
\label{gb4bD4Btwo1}
0=&\, \left[ \frac{1}{4\kappa^2} R + \frac{1}{2}
\left\{
 - \frac{1}{2} A \partial_\rho \phi \partial^\rho \phi
 - B \partial_\rho \phi \partial^\rho \chi
 - \frac{1}{2} C \partial_\rho \chi \partial^\rho \chi - V \right\}
 - 4 \left( \nabla_\rho \nabla_\sigma \xi \right) R^{\rho\sigma} \right] h_{\mu\nu} \nonumber \\
&\, + \bigg[ \frac{1}{4} g_{\mu\nu} \left\{
 - A \partial^\tau \phi \partial^\eta \phi
 - B \left(\partial^\tau \phi \partial^\eta \chi + \partial^\eta \phi \partial^\tau \chi \right)
 - C \partial^\tau \chi \partial^\eta \chi \right\} \nonumber \\
&\, - 2 g_{\mu\nu} \left( \nabla^\tau \nabla^\eta \xi\right)R
 - 4 \left( \nabla^\tau \nabla_\mu \xi\right)R_\nu^{\ \eta} - 4 \left( \nabla^\tau \nabla_\nu \xi\right)R_\mu^{\ \eta}
+ 4 \left( \nabla^\tau \nabla^\eta \xi \right)R_{\mu\nu} \nonumber \\
&\, + 4g_{\mu\nu} \left( \nabla^\tau \nabla_\sigma \xi \right)
R^{\eta\sigma} + 4g_{\mu\nu} \left( \nabla_{\rho} \nabla^\tau \xi
\right) R^{\rho\eta}
 - 4 \left(\nabla^\tau \nabla^\sigma \xi \right) R_{\mu\ \, \nu\sigma}^{\ \, \eta}
 - 4 \left(\nabla^\rho \nabla^\tau \xi \right) R_{\mu\rho\nu}^{\ \ \ \ \eta}
\bigg\} h_{\tau\eta} \nonumber \\
&\, + \frac{1}{2}\left\{ 2 \delta_\mu^{\ \eta} \delta_\nu^{\
\zeta} \left( \nabla_\kappa \xi \right)R
 - 4 \delta_\rho^{\ \eta} \delta_\mu^{\ \zeta} \left( \nabla_\kappa \xi \right)R_\nu^{\ \rho}
 - 4 \delta_\rho^{\ \eta} \delta_\nu^{\ \zeta} \left( \nabla_\kappa \xi \right)R_\mu^{\ \rho} \right. \nonumber \\
&\, \left. + 4g_{\mu\nu} \delta_\rho^{\ \eta} \delta_\sigma^{\
\zeta} \left( \nabla_\kappa \xi \right) R^{\rho\sigma}
 - 4 g^{\rho\eta} g^{\sigma\zeta} \left( \nabla_\kappa \xi \right) R_{\mu\rho\nu\sigma}
\right\} g^{\kappa\lambda}\left( \nabla_\eta h_{\zeta\lambda} + \nabla_\zeta h_{\eta\lambda} - \nabla_\lambda h_{\eta\zeta} \right) \nonumber \\
&\, - \left\{ \frac{1}{4\kappa^2} g_{\mu\nu} - 2 \left( \nabla_\mu
\nabla_\nu \xi \right) + 2 g_{\mu\nu} \left( \nabla^2\xi \right)
\right\}
R^{\mu\nu} h_{\mu\nu} \nonumber \\
&\, + \frac{1}{2}\left\{ \left( - \frac{1}{2\kappa^2} - 4 \nabla^2
\xi \right) \delta^\tau_{\ \mu} \delta^\eta_{\ \nu} + 4 \left(
\nabla_\rho \nabla_\mu \xi \right) \delta^\eta_{\ \nu}
g^{\rho\tau} + 4 \left( \nabla_\rho \nabla_\nu \xi \right)
\delta^\tau_{\ \mu} g^{\rho\eta}
 - 4g_{\mu\nu} \nabla^\tau \nabla^\eta \xi \right\} \nonumber \\
&\, \qquad \times \left\{ - \nabla^2 h_{\tau\eta} - 2R^{\lambda\
\phi}_{\ \eta\ \tau}h_{\lambda\phi}
+ R^\phi_{\ \tau}h_{\phi\eta} + R^\phi_{\ \tau}h_{\phi\eta} \right\} \nonumber \\
&\, + 2 \left(\nabla^\rho \nabla^\sigma \xi \right) \left\{
\nabla_\nu \nabla_\rho h_{\sigma\mu}
 - \nabla_\nu \nabla_\mu h_{\sigma\rho}
 - \nabla_\sigma \nabla_\rho h_{\nu\mu}
 + \nabla_\sigma \nabla_\mu h_{\nu\rho}
+ h_{\mu\phi} R^\phi_{\ \rho\nu\sigma}
 - h_{\rho\phi} R^\phi_{\ \mu\nu\sigma} \right\} \nonumber \\
&\, + \frac{1}{2}\frac{\partial T_{\mathrm{matter}\,
\mu\nu}}{\partial g_{\tau\eta}}h_{\tau\eta} \, .
\end{align}
In order to investigate the propagating speed $c_\mathrm{GW}$ of
the gravitational wave $h_{\mu\nu}$, we check the parts including
the second derivatives of $h_{\mu\nu}$ and we reobtain
(\ref{condition}), although $\xi$ is now a function of the two
scalar fields $\phi$ and $\chi$, $\xi=\xi(\phi,\chi)$. Therefore
even in the case of Einstein-Gauss-Bonnet gravity coupled with two
scalars, it is impossible to obtain a model for which the
propagating speed of the gravitational wave coincides with that of
light.

\subsection{Propagating Speed of Gravitational Wave Inside Stellar Objects}\label{Sec6}

We have shown that the propagating speed of gravitational waves in
a non-trivial spherically symmetric spacetime is always different
from the speed of light, even in the Einstein-Gauss-Bonnet gravity
with two scalar fields (\ref{I8}). In this subsection, we estimate
the shift of the gravitational wave speed in stellar objects in
the context of Einstein-Gauss-Bonnet gravity with only one scalar
field $\phi$ in (\ref{I8one}).

For the metric given by Eq.~(\ref{GBiv}), the $(t,t)$-, $(r,r)$-,
the angular components of Eq.~(\ref{gb4bD4}) and the equation for
the scalar field $\phi$, have the following forms,
\begin{align}
\label{Eq2tt}
 - 4r^2 \e^{2\lambda} \kappa^2 \rho =&\,- 16 \left(1 - \e^{-2\lambda}\right) \xi'' - 4 \left\{ - 4\left( 1-3\e^{-2\lambda} \right)\xi' + r \right\} \lambda'
+2 +r^2 \phi'^2 + 2 \e^{2\lambda} \left( V r^2 -1 \right) \,, \\
\label{Eq2rr}
4 r^2 \e^{2\lambda} \kappa^2 p =&\, 4\left\{ - 4 \left(1 -3\e^{-2\lambda} \right) \xi' +r \right\} \nu' + 2 - r^2 \phi'^2 -2  \e^{-2\lambda} + 2 \e^{2\lambda} V r^2 \,, \\
\label{Eq2pp}
8r \e^{2\lambda} \kappa^2 p =&\, 2 \left(r + 8 \xi' \e^{-2\lambda} \right) \left( \nu'' + {\nu'}^2 \right) + 16 \xi'' \nu' \e^{-2\lambda}
+ \left\{ -2 \left( r + 24 \xi' \e^{-2\lambda} \right) \lambda' + 2 \right\} \nu' \nonumber \\
&\, -2 \lambda' + r\left( {\phi'}^2 + 2 \e^{2\lambda} V \right) \, , \\
\label{Eqphi2}
0=&\, - 8 \xi' \left(\e^{-2\lambda}-1 \right) \left( \nu'' + 2{\nu'}^2 \right) +\phi' \phi'' r^2
 - 8 \nu' \xi' \left\{ \nu' \left(1-\e^{-2\lambda} \right) - \lambda' \left( 3\e^{-2\lambda} - 1 \right) \right\} \nonumber \\
&\, +r \left( \nu' r + 2 - \lambda' r \right)\phi'^2 - \e^{2\lambda} V' r^2 \,.
\end{align}
Here $\rho$ is the energy density and $p$ is the pressure of matter, which we
assume to be a perfect fluid { and} satisfies an equation of state, $p=p\left(\rho\right)$.
The energy density $\rho$ and the pressure $p$ satisfy the following conservation law,
\begin{align}
\label{FRN2}
0 = \nabla^\mu T_{\mu r} =\nu' \left( \rho + p \right) + \frac{dp}{dr} \, .
\end{align}
The conservation law is also derived from Eqs.~(\ref{Eq2tt}),
(\ref{Eq2rr}), (\ref{Eq2pp}), and (\ref{Eqphi2}).
Here we have assumed that $\rho$ and $p$ depend only on the radial coordinate $r$.
Other components of the conservation law are trivially
satisfied. If the equation of state $\rho=\rho(p)$ is given,
Eq.~(\ref{FRN2}) can be integrated as follows,
\begin{align}
\label{FRN3}
\nu = - \int^r dr \frac{\frac{dp}{dr}}{\rho + p}
= - \int^{p(r)}\frac{dp}{\rho(p) + p} \, .
\end{align}
Because Eq.~(\ref{FRN2}) and therefore (\ref{FRN3}), can be
obtained from Eqs.~(\ref{Eq2tt}), (\ref{Eq2rr}), (\ref{Eq2pp}),
and (\ref{Eqphi2}), as long as we use (\ref{FRN3}), we forget one
equation in Eqs.~(\ref{Eq2tt}), (\ref{Eq2rr}), (\ref{Eq2pp}), and (\ref{Eqphi2}).
In the following, we do not use Eq.~(\ref{Eqphi2}).
Inside the compact stellar object, we can use
Eq.~(\ref{FRN3}) but outside the stellar object, we cannot use
Eq.~(\ref{FRN3}). Instead of using Eq.~(\ref{FRN3}), we may assume
the profile of $\nu=\nu(r)$ so that $\nu(r)$ and $\nu'(r)$ are
continuous at the surface of the compact stellar object.

By combining Eq.~(\ref{Eq2tt}) with Eq.~(\ref{Eq2rr}), we obtain,
\begin{align}
\label{V2}
V =&\, \kappa^2 \left( - \rho + p \right) + \frac{\e^{-2\lambda}}{r^2}\left\{ - 4 \left( \e^{-2\lambda}-1 \right) \xi'' - 4 \left( 1-3\e^{-2\lambda} \right)  (\lambda' -\nu')
 \xi' +\e^{2\lambda} - 1 \right\} +\frac{\e^{-2\lambda}}{r} \left( \lambda' -\nu' \right) \, , \\
\label{xi2}
\phi' =&\, \pm \sqrt{ - 2 \e^{2\lambda} \kappa^2 \left( \rho + p \right) - \frac{8}{r^2} \left\{ \left( \e^{-2\lambda}-1 \right) \xi''
+ \left( 1-3\e^{-2\lambda} \right) \left( \lambda' +\nu' \right) \xi'\right\}
+\frac{2}{r} \left( \lambda' +\nu' \right) } \, .
\end{align}
Furthermore, the combination of Eq.~(\ref{Eq2tt}) and Eq.~(\ref{Eq2pp}) gives,
\begin{align}
\label{f2}
0 =&\, - 8\, \left\{ \e^{-2\lambda} \left( \nu' r - 1 \right) +1 \right\} \xi'' - 8 \e^{-2\lambda} \left\{ r \left( \nu'' + {\nu'}^2 -3  \nu' \lambda' \right)
+ \lambda' \left( 3 -\e^{2\lambda} \right)  \right\} \xi' \nonumber \\
&\, -r^2 \left( \nu'' +{\nu'}^2 -\nu' \lambda' \right) -2r \left( \nu' +\lambda' \right) - \e^{2\lambda} +1 - 2 \kappa^2 r^2 \e^{2\lambda} \left( \rho + p \right)\, ,
\end{align}
which can be regarded as a differential equation for $\xi'$ and
therefore for $\xi$ if $\nu=\nu(r)$, $\lambda = \lambda(r)$,
$\rho=\rho(r)$, and $p=p(r)$ given, the solution is,
\begin{align}
\label{f3}
\xi(r) = &\, - \frac{1}{8}\int \left[ \int \frac{
\e^{2\lambda} \left\{ \e^{2\lambda} + r^2 \left( \nu'' +{\nu'}^2 -\nu' \lambda' \right) +r (\nu' +\lambda') -1 - 2 \kappa^2 r^2 \e^{2\lambda} \left( \rho + p \right)\right\} }
{U \left(\nu' r - 1 + \e^{2\lambda} \right)} dr + c_1 \right] U dr +c_2
 \, , \nonumber \\
U(r) \equiv&\, \exp \left\{ -\int \frac{  r \left( \nu'' +
{\nu'}^2 \right) + \lambda' \left( 3 - \e^{2\lambda} - 3 \nu' r
\right)} {\nu' r - 1 + \e^{2\lambda}} dr \right\} \, ,
\end{align}
where $c_1$ and $c_2$ are integration constants.
We may properly assume the profile of $\nu=\nu(r)$ and $\lambda = \lambda(r)$.
Therefore, by using (\ref{f3}), we find the $r$-dependence of
$\xi$, $\xi=\xi(r)$ and by using Eqs.~(\ref{V2}) and (\ref{xi2}),
we find the $r$ dependencies of $V$ and $\phi$, $V=V(r)$ and $\phi=\phi(r)$.
By solving $\phi=\phi(r)$ with respect to $r$,
$r=r(\phi)$, we find $\xi$ and $V$ as functions of $\phi$,
$\xi(\phi)=\xi\left( r \left( \phi \right) \right)$,
$V(\phi)=V\left( r \left( \phi \right) \right)$ which realize the
model which has a solution given by $\nu=\nu(r)$ and $\lambda = \lambda(r)$.
We should note, however, the expression of $\phi$ in
(\ref{xi2}) gives the following constraint,
\begin{align}
\label{cons3}
 - 2 \e^{2\lambda}\kappa^2 \left( \rho + p \right) - \frac{8}{r^2} \left\{ \left( \e^{-2\lambda}-1 \right) \xi''
 + \left( 1-3\e^{-2\lambda} \right) \left( \lambda' +\nu' \right) \xi'\right\}
+\frac{2}{r} \left( \lambda' +\nu' \right) \geq 0 \, ,
\end{align}
so that the ghost could be avoided. If Eq.~(\ref{cons3}) is not
satisfied, the scalar field $\phi$ becomes pure imaginary.
We may define a new real scalar field $\zeta$ by $\phi=i\zeta$
$\left(i^2=-1\right)$ but because the coefficient in front of the
kinetic term of $\zeta$ becomes negative, $\zeta$ is rendered a ghost.
The existence of the ghost generates the negative norm states in the quantum theory and
therefore the theory becomes inconsistent.

When we consider compact stellar objects like neutron stars, we
often consider the following equation of state,
\begin{enumerate}
\item Energy-polytrope
\begin{align}
\label{polytrope}
p = K \rho^{1 + \frac{1}{n}}\,,
\end{align}
with constants $K$ and $n$. It is known that for the neutron
stars, $n$ could take the value $0.5\leq n \leq 1$.
\item Mass-polytrope
\begin{align}
\label{MassPolytropicEOS}
\rho = \rho_{m} + N p \, ,\qquad \qquad p = K_m \rho_m^{1+\frac{1}{n_{m}}} \, ,
\end{align}
where $\rho_{m}$ is rest mass energy density and $K_{m}$, $N$ are constants.
\end{enumerate}
Now let us study the case of the energy-polytrope
(\ref{polytrope}), in detail, in which we can rewrite the equation of state as follows,
\begin{align}
\label{polytrope2}
\rho = \tilde K p^{({1 + \frac{1}{\tilde n}})}\, , \quad
\tilde K \equiv K^{-\frac{1}{1+\frac{1}{n}}} \, , \quad
\tilde n \equiv \frac{1}{\frac{1}{1+\frac{1}{n}} - 1}
= - 1 - n \, .
\end{align}
For the energy-polytrope, Eq.~(\ref{FRN3}) takes the following form,
\begin{align}
\label{FRN3p1B}
\nu = - \int^{p(r)}\frac{dp}{\tilde K p^{1 + \frac{1}{\tilde n}} + p}
= \frac{c}{2} + \tilde n \ln \left(1+{\tilde K}^{-1}p^{-\frac{1}{\tilde n}} \right)
= \frac{c}{2} - \left(1+n\right) \ln \left(1+ K \rho^\frac{1}{n} \right) \, ,
\end{align}
where $c$ is an integration constant. Similarly, in the case of
mass-polytrope (\ref{MassPolytropicEOS}), we obtain,
\begin{align}
\label{masspolytope} \nu = \frac{\tilde c}{2} + \ln \left( 1 - K_m
\rho^\frac{1}{n_m}\right) \, ,
\end{align}
where $\tilde c$ is an integration constant.

We now consider the energy-polytrope in Eq.~(\ref{FRN3}) and
investigate the behavior of the solution in the region around the center of the stellar object.
In order to avoid a conical singularity at the center of the stellar object, we require the following
behavior of $\rho$ near the center of the star,
\begin{align}
\label{azrho} \rho \sim\rho_0 + \rho_2 r^2 \, , \quad \lambda=
\lambda_2 r^2 \, ,
\end{align}
where $\rho_0$, $\rho_2$, and $\lambda_2$ are constants.
We should note that when $r\to 0$, we need to require $\lambda, \lambda' \to 0$ in order to avoid the conical singularity.
Then Eq.~(\ref{polytrope}) and Eq.~(\ref{FRN3p1B}) gives,
\begin{align}
\label{pandnu}
p \sim&\, p_0 + p_2 r^2\, , \quad p_0 \equiv K {\rho_0}^{1+\frac{1}{n}} \, , \quad p_2 \equiv K {\rho_0}^{1+\frac{1}{n}} \left(1+\frac{1}{n} \right) \frac{\rho_2}{\rho_0} \, , \nonumber \\
\nu \sim&\, \nu_0 + \nu_2 r^2 \, , \quad \nu_0 \equiv \frac{c}{2} - \left( 1 + n \right) \ln \left( 1 + K {\rho_0}^\frac{1}{n} \right) \, , \quad
\nu_2 \equiv - \left( 1 + \frac{1}{n} \right) \frac{K {\rho_0}^{\frac{1}{n}-1}\rho_2}{1 + K {\rho_0}^\frac{1}{n}}  \, .
\end{align}
Therefore by using (\ref{f3}), we obtain,
\begin{align}
\label{f3R1}
\xi'(r) = \xi_1 + 2 \xi_2 r \, , \quad \xi_1 \equiv \frac{c_1}{8} \, , \quad
\xi_2 \equiv - \frac{2 \left( \nu_2 + \lambda_2 \right) + \kappa^2 \left( \rho_0 + K {\rho_0}^{1+\frac{1}{n}} \right)}{8 \left( \nu_2 + \lambda_2 \right)} \, .
\end{align}
In order to avoid the conical singularity, we need to require $\xi_1=0$.

For simplicity, we assume,
\begin{align}
\label{hij} h_{ij} = \frac{\mathrm{Re} \left( \e^{-i\omega t + i k
r} \right) h_{ij}^{(0)}}{r} \quad \left( i,j =\theta, \phi, \quad
\sum_i h_{\ \ i}^{(0)\ i}=0 \right)\, , \quad \mbox{other
components}=0\, ,
\end{align}
where $h_{ij}^{(0)}$'s are constants corresponding to the polarization.
At the center of the stellar object, by using (\ref{nablaxi}), we find,
\begin{align}
\label{nablaxis}
\nabla_t \nabla_t \xi =&\, \nabla_i \nabla_j \xi = \nabla_r \nabla_t \xi = \nabla_t \nabla_r \xi = \nabla_t \nabla_i \xi
= \nabla_i \nabla_t \xi = \nabla_r \nabla_i \xi = \nabla_i \nabla_r \xi = 0 \, ,
\nonumber \\
\quad \nabla_r \nabla_r \xi =&\, 2 \xi_2\, , \quad \nabla^i \nabla^j \xi = 2 \xi_2 {\tilde g}^{ij} \, , \quad \nabla^2 \xi = 6 \xi_2 \, .
\end{align}
Then when the energy of the gravitational wave is large, by using
(\ref{second}), we find the following dispersion relation,
\begin{align}
\label{dispersion}
0= - \frac{1}{2} \left( \frac{1}{2\kappa^2} + 24 \xi_2 \right) \e^{-2\nu_0} \omega^2
+ \frac{1}{2} \left( \frac{1}{2\kappa^2} + 16 \xi_2 \right) \xi_2 k^2 \, ,
\end{align}
which indicates that the propagating speed $c_\mathrm{GW}$ of the
gravitational wave is given by,
\begin{align}
\label{cGW}
{c_\mathrm{GW}}^2 = \left( \frac{1 + 32 \kappa^2 \xi_2 }{1 + 48 \kappa^2 \xi_2} \right) c^2 \, .
\end{align}
We should note that the speed of light $c$ is now given by $c^2 = \e^{2 \nu_0}$.
If $\left| \kappa^2 \xi_2 \right| \ll 1$, then Eq.~(\ref{cGW}) is approximated as,
\begin{align}
\label{cGW}
{c_\mathrm{GW}}^2 \sim \left( 1 - 16 \kappa^2 \xi_2 \right) c^2 \, .
\end{align}
Therefore if $\xi_2>0$ $\left(\xi_2<0\right)$, the propagating
speed of the gravitational wave becomes larger (smaller) than that of light.

The GW170817 event in (\ref{GWp9}) gives a strong constraint on
the parameter $\xi_2$ as follows,
\begin{align}
\label{GWp9stlr}
\left| 16 \kappa^2 \xi_2 \right| < 6 \times 10^{-15}\, .
\end{align}
In the limit $\xi_2\to 0$, $\xi$ becomes almost constant near the
center of the stellar objects and Eq.~(\ref{f3R1}) indicates that,
\begin{align}
\label{f3R1stlr}
0= 2 \left( \nu_2 + \lambda_2 \right) + \kappa^2 \left( \rho_0 + K {\rho_0}^{1+\frac{1}{n}} \right) \, ,
\end{align}
which is consistent with Einstein's gravity with $\xi(\phi)=0$.
In fact, Eq.~(\ref{f3R1stlr}) is obtained from Eq.~(\ref{f2}) by putting $\xi(\phi)=0$ and by using
(\ref{azrho}) and (\ref{pandnu}) at the center.

\section{Summary and Discussion}\label{Sec9}

In this paper, we have investigated the propagating speed of the
gravitational waves in the spherically symmetric spacetime and
cosmological spacetimes of the FLRW form, which are solutions of
Einstein-Gauss-Bonnet gravity. We have found, that there is no
possibility that the speed could coincide with that of light in
spherically symmetric backgrounds. We estimated the shift of the
propagating speed inside stellar objects and in several epochs
like during the inflation, the end of inflation, the reheating,
and late time era in the framework of the Einstein-Gauss-Bonnet
gravity coupled with one scalar field. In order not to conflict
with the GW170817 observations~\cite{TheLIGOScientific:2017qsa,
Monitor:2017mdv, GBM:2017lvd}, we have proposed a scenario that
Einstein-Gauss-Bonnet gravity reduces to the standard
scalar-tensor theory in late times by requiring that the
Gauss-Bonnet coupling $\xi(\phi)$ of the scalar field in the
action (\ref{I8one}) and also the scalar field $\phi$ goes to a
constant in the late time era, although the Gauss-Bonnet coupling
may play important and non-trivial roles in the early Universe. We
constructed a rather realistic model that could satisfy the above
requirement. An interesting point could be that the model would
describe both the inflationary era in the early Universe and the
accelerating expansion of the late Universe without introducing
parameters with so large hierarchy.

What could happen when the propagating speed of the gravitational
wave is different from that of light in the early Universe? For
the fixed frequency, the wavelength becomes longer (shorter) if
the speed is larger (smaller) than the light speed. Usually, the
wave with a longer wavelength generates higher output. Therefore
if the speed is larger (smaller), the primordial gravitational
wave becomes more (less) abundant. Another point is a cosmological
horizon. For gravity-related fluctuations, the cosmological
horizon becomes larger (smaller) if the speed of the gravitational
wave becomes larger (smaller) than those of other modes including
light and scalar fields. Therefore the tensor mode and so-called
B-mode polarization could be affected.

In the case of the stellar object, we have estimated the
propagating speed of the gravitational wave at the center of the
stellar object, in order to avoid any ambiguities. If the
Gauss-Bonnet coupling $\xi(\phi)$ becomes almost constant, most of
the gravitational waves propagate at the speed of light. Such a
behavior $\xi(\phi)$ strongly depends on the details of the model.
There could be, however, the model where $\xi(\phi)$ could depend
on the coordinates even outside of the stellar objects, especially
in the case of compact stars like neutron stars. Furthermore,
there might be a small portion of the gravitational wave that goes
through inside the stellar objects. If the propagating speed of
the gravitational wave is larger than that of light, there might
be a small signal before the main part of the gravitational wave
is observed. If the propagating speed of the gravitational wave is
larger than that of light, however, the causality could be
violated and therefore it might be prohibited. In this case, there
could be some interesting phenomena. For example, some information
inside the black hole horizon, which is a null surface, might go
through outside the horizon, which may solve the problem of the
information paradox of the black hole. Finally, let us note that
Einstein-Gauss-Bonnet gravity may lead to finite-time future
singularity (for a review see \cite{deHaro:2023lbq}) and it would
be of interest to study the gravitational wave speed in
Einstein-Gauss-Bonnet theory when the Universe reaches a future
finite-time singularity.

\section*{Acknowledgements}

This work was partially supported by MICINN (Spain), project
PID2019-104397GB-I00  and by the program Unidad de Excelencia
Maria de Maeztu CEX2020-001058-M, Spain (S.D.O). S.N. was partly
supported by MdM Core visiting professorship at ICE-CSIC,
Barcelona.



\begin{thebibliography}{99}

\bibitem{CMB-S4:2016ple}
K.~N.~Abazajian \textit{et al.} [CMB-S4],
[arXiv:1610.02743 [astro-ph.CO]].



\bibitem{SimonsObservatory:2019qwx}
M.~H.~Abitbol \textit{et al.} [Simons Observatory],
Bull. Am. Astron. Soc. \textbf{51} (2019), 147 [arXiv:1907.08284
[astro-ph.IM]].




\bibitem{Hild:2010id}
S.~Hild, M.~Abernathy, F.~Acernese, P.~Amaro-Seoane, N.~Andersson,
K.~Arun, F.~Barone, B.~Barr, M.~Barsuglia and M.~Beker, \textit{et
al.}
Class. Quant. Grav. \textbf{28} (2011), 094013
doi:10.1088/0264-9381/28/9/094013 [arXiv:1012.0908 [gr-qc]].




\bibitem{Baker:2019nia}
J.~Baker, J.~Bellovary, P.~L.~Bender, E.~Berti, R.~Caldwell,
J.~Camp, J.~W.~Conklin, N.~Cornish, C.~Cutler and R.~DeRosa,
\textit{et al.}
[arXiv:1907.06482 [astro-ph.IM]].


\bibitem{Smith:2019wny}
T.~L.~Smith and R.~Caldwell,
Phys. Rev. D \textbf{100} (2019) no.10, 104055
doi:10.1103/PhysRevD.100.104055 [arXiv:1908.00546 [astro-ph.CO]].


\bibitem{Crowder:2005nr}
J.~Crowder and N.~J.~Cornish,
Phys. Rev. D \textbf{72} (2005), 083005
doi:10.1103/PhysRevD.72.083005 [arXiv:gr-qc/0506015 [gr-qc]].


\bibitem{Smith:2016jqs}
T.~L.~Smith and R.~Caldwell,
Phys. Rev. D \textbf{95} (2017) no.4, 044036
doi:10.1103/PhysRevD.95.044036 [arXiv:1609.05901 [gr-qc]].



\bibitem{Seto:2001qf}
N.~Seto, S.~Kawamura and T.~Nakamura,
Phys. Rev. Lett. \textbf{87} (2001), 221103
doi:10.1103/PhysRevLett.87.221103 [arXiv:astro-ph/0108011
[astro-ph]].


\bibitem{Kawamura:2020pcg}
S.~Kawamura, M.~Ando, N.~Seto, S.~Sato, M.~Musha, I.~Kawano,
J.~Yokoyama, T.~Tanaka, K.~Ioka and T.~Akutsu, \textit{et al.}
[arXiv:2006.13545 [gr-qc]].



\bibitem{Bull:2018lat}
A.~Weltman, P.~Bull, S.~Camera, K.~Kelley, H.~Padmanabhan,
J.~Pritchard, A.~Raccanelli, S.~Riemer-S\o{}rensen, L.~Shao and
S.~Andrianomena, \textit{et al.}
Publ. Astron. Soc. Austral. \textbf{37} (2020), e002
doi:10.1017/pasa.2019.42 [arXiv:1810.02680 [astro-ph.CO]].




\bibitem{LISACosmologyWorkingGroup:2022jok}
P.~Auclair \textit{et al.} [LISA Cosmology Working Group],
[arXiv:2204.05434 [astro-ph.CO]].




\bibitem{NANOGrav:2023gor}
G.~Agazie \textit{et al.} [NANOGrav],
Astrophys. J. Lett. \textbf{951} (2023) no.1, L8
doi:10.3847/2041-8213/acdac6 [arXiv:2306.16213 [astro-ph.HE]].



\bibitem{sunnynew}
S.~Vagnozzi,
JHEAp \textbf{39} (2023), 81-98 doi:10.1016/j.jheap.2023.07.001
[arXiv:2306.16912 [astro-ph.CO]].


\bibitem{Nojiri:2010wj}
S.~Nojiri and S.~D.~Odintsov,
Phys. Rept. \textbf{505} (2011), 59-144
doi:10.1016/j.physrep.2011.04.001
[arXiv:1011.0544 [gr-qc]].

\bibitem{Nojiri:2017ncd}
S.~Nojiri, S.~D.~Odintsov and V.~K.~Oikonomou,
Phys. Rept. \textbf{692} (2017), 1-104
doi:10.1016/j.physrep.2017.06.001
[arXiv:1705.11098 [gr-qc]].

\bibitem{Hwang:2005hb}
 J.~c.~Hwang and H.~Noh,
 Phys.\ Rev.\ D {\bf 71} (2005) 063536
 doi:10.1103/PhysRevD.71.063536
 [gr-qc/0412126].

\bibitem{Nojiri:2006je}
 S.~Nojiri, S.~D.~Odintsov and M.~Sami,
 Phys.\ Rev.\ D {\bf 74} (2006) 046004
 doi:10.1103/PhysRevD.74.046004
 [hep-th/0605039].

\bibitem{Nojiri:2005vv}
 S.~Nojiri, S.~D.~Odintsov and M.~Sasaki,
 Phys.\ Rev.\ D {\bf 71} (2005) 123509
 doi:10.1103/PhysRevD.71.123509
 [hep-th/0504052].

\bibitem{Satoh:2007gn}
 M.~Satoh, S.~Kanno and J.~Soda,
 Phys.\ Rev.\ D {\bf 77} (2008) 023526
 doi:10.1103/PhysRevD.77.023526
 [arXiv:0706.3585 [astro-ph]].

\bibitem{Yi:2018gse}
 Z.~Yi, Y.~Gong and M.~Sabir,
 Phys.\ Rev.\ D {\bf 98} (2018) no.8, 083521
 doi:10.1103/PhysRevD.98.083521
 [arXiv:1804.09116 [gr-qc]].

\bibitem{Guo:2009uk}
 Z.~K.~Guo and D.~J.~Schwarz,
 Phys.\ Rev.\ D {\bf 80} (2009) 063523
 doi:10.1103/PhysRevD.80.063523
 [arXiv:0907.0427 [hep-th]].

\bibitem{Jiang:2013gza}
 P.~X.~Jiang, J.~W.~Hu and Z.~K.~Guo,
 Phys.\ Rev.\ D {\bf 88} (2013) 123508
 doi:10.1103/PhysRevD.88.123508
 [arXiv:1310.5579 [hep-th]].

\bibitem{Kanti:2015pda}
 P.~Kanti, R.~Gannouji and N.~Dadhich,
 Phys.\ Rev.\ D {\bf 92} (2015) no.4, 041302
 doi:10.1103/PhysRevD.92.041302
 [arXiv:1503.01579 [hep-th]].

\bibitem{vandeBruck:2017voa}
 C.~van de Bruck, K.~Dimopoulos, C.~Longden and C.~Owen,
 arXiv:1707.06839 [astro-ph.CO].

\bibitem{Kanti:1998jd}
 P.~Kanti, J.~Rizos and K.~Tamvakis,
 Phys.\ Rev.\ D {\bf 59} (1999) 083512
 doi:10.1103/PhysRevD.59.083512
 [gr-qc/9806085].

\bibitem{Pozdeeva:2020apf}
E.~O.~Pozdeeva, M.~R.~Gangopadhyay, M.~Sami, A.~V.~Toporensky and
S.~Y.~Vernov,
Phys. Rev. D \textbf{102} (2020) no.4, 043525
doi:10.1103/PhysRevD.102.043525 [arXiv:2006.08027 [gr-qc]].

\bibitem{Pozdeeva:2021iwc}
E.~O.~Pozdeeva and S.~Y.~Vernov,
Eur. Phys. J. C \textbf{81} (2021) no.7, 633
doi:10.1140/epjc/s10052-021-09435-8 [arXiv:2104.04995 [gr-qc]].

\bibitem{Koh:2014bka}
S.~Koh, B.~H.~Lee, W.~Lee and G.~Tumurtushaa,
Phys. Rev. D \textbf{90} (2014) no.6, 063527
doi:10.1103/PhysRevD.90.063527 [arXiv:1404.6096 [gr-qc]].

\bibitem{Bayarsaikhan:2020jww}
B.~Bayarsaikhan, S.~Koh, E.~Tsedenbaljir and G.~Tumurtushaa,
JCAP \textbf{11} (2020), 057 doi:10.1088/1475-7516/2020/11/057
[arXiv:2005.11171 [gr-qc]].

\bibitem{DeLaurentis:2015fea}
 M.~De Laurentis, M.~Paolella and S.~Capozziello,
 Phys.\ Rev.\ D {\bf 91} (2015) no.8, 083531
 doi:10.1103/PhysRevD.91.083531
 [arXiv:1503.04659 [gr-qc]].

\bibitem{Chervon:2019sey}
 Scalar Field Cosmology, S.~Chervon, I.~Fomin, V.~Yurov and
 A.~Yurov, World Scientific 2019, \\ doi:10.1142/11405

\bibitem{Nozari:2017rta}
 K.~Nozari and N.~Rashidi,
 Phys.\ Rev.\ D {\bf 95} (2017) no.12, 123518
 doi:10.1103/PhysRevD.95.123518
 [arXiv:1705.02617 [astro-ph.CO]].

\bibitem{Odintsov:2018zhw}
 S.~D.~Odintsov and V.~K.~Oikonomou,
 Phys.\ Rev.\ D {\bf 98} (2018) no.4, 044039
 doi:10.1103/PhysRevD.98.044039
 [arXiv:1808.05045 [gr-qc]].

 \bibitem{Kawai:1998ab}
 S.~Kawai, M.~a.~Sakagami and J.~Soda,
 Phys.\ Lett.\ B {\bf 437}, 284 (1998)
 doi:10.1016/S0370-2693(98)00925-3
 [gr-qc/9802033].

\bibitem{Yi:2018dhl}
 Z.~Yi and Y.~Gong,
 Universe {\bf 5} (2019) no.9, 200
 doi:10.3390/universe5090200
 [arXiv:1811.01625 [gr-qc]].

\bibitem{vandeBruck:2016xvt}
 C.~van de Bruck, K.~Dimopoulos and C.~Longden,
 Phys.\ Rev.\ D {\bf 94} (2016) no.2, 023506
 doi:10.1103/PhysRevD.94.023506
 [arXiv:1605.06350 [astro-ph.CO]].

\bibitem{Kleihaus:2019rbg}
B.~Kleihaus, J.~Kunz and P.~Kanti,
Phys. Lett. B \textbf{804} (2020), 135401
doi:10.1016/j.physletb.2020.135401 [arXiv:1910.02121 [gr-qc]].


\bibitem{Bakopoulos:2019tvc}
 A.~Bakopoulos, P.~Kanti and N.~Pappas,
 Phys.\ Rev.\ D {\bf 101} (2020) no.4, 044026
 doi:10.1103/PhysRevD.101.044026
 [arXiv:1910.14637 [hep-th]].

\bibitem{Maeda:2011zn}
 K.~i.~Maeda, N.~Ohta and R.~Wakebe,
 Eur.\ Phys.\ J.\ C {\bf 72} (2012) 1949
 doi:10.1140/epjc/s10052-012-1949-6
 [arXiv:1111.3251 [hep-th]].


\bibitem{Bakopoulos:2020dfg}
A.~Bakopoulos, P.~Kanti and N.~Pappas,
Phys. Rev. D \textbf{101} (2020) no.8, 084059
doi:10.1103/PhysRevD.101.084059 [arXiv:2003.02473 [hep-th]].

\bibitem{Ai:2020peo}
W.~Y.~Ai,
Commun. Theor. Phys. \textbf{72} (2020) no.9, 095402
doi:10.1088/1572-9494/aba242 [arXiv:2004.02858 [gr-qc]].

\bibitem{Odintsov:2020xji}
S.~D.~Odintsov, V.~K.~Oikonomou and F.~P.~Fronimos,
Annals Phys. \textbf{420} (2020), 168250
doi:10.1016/j.aop.2020.168250 [arXiv:2007.02309 [gr-qc]].

\bibitem{Oikonomou:2020sij}
V.~K.~Oikonomou and F.~P.~Fronimos,
Class. Quant. Grav. \textbf{38} (2021) no.3, 035013
doi:10.1088/1361-6382/abce47 [arXiv:2006.05512 [gr-qc]].

\bibitem{Odintsov:2020zkl}
S.~D.~Odintsov and V.~K.~Oikonomou,
Phys. Lett. B \textbf{805} (2020), 135437
doi:10.1016/j.physletb.2020.135437 [arXiv:2004.00479 [gr-qc]].

\bibitem{Odintsov:2020mkz}
S.~D.~Odintsov, V.~K.~Oikonomou, F.~P.~Fronimos and
S.~A.~Venikoudis,
Phys. Dark Univ. \textbf{30} (2020), 100718
doi:10.1016/j.dark.2020.100718 [arXiv:2009.06113 [gr-qc]].

\bibitem{Venikoudis:2021irr}
S.~A.~Venikoudis and F.~P.~Fronimos,
Eur. Phys. J. Plus \textbf{136} (2021) no.3, 308
doi:10.1140/epjp/s13360-021-01298-y [arXiv:2103.01875 [gr-qc]].

\bibitem{Kong:2021qiu}
S.~B.~Kong, H.~Abdusattar, Y.~Yin and Y.~P.~Hu,
[arXiv:2108.09411 [gr-qc]].

\bibitem{Easther:1996yd}
 R.~Easther and K.~i.~Maeda,
 Phys.\ Rev.\ D {\bf 54} (1996) 7252
 doi:10.1103/PhysRevD.54.7252
 [hep-th/9605173].

\bibitem{Antoniadis:1993jc}
 I.~Antoniadis, J.~Rizos and K.~Tamvakis,
 Nucl.\ Phys.\ B {\bf 415} (1994) 497
 doi:10.1016/0550-3213(94)90120-1
 [hep-th/9305025].

\bibitem{Antoniadis:1990uu}
I.~Antoniadis, C.~Bachas, J.~R.~Ellis and D.~V.~Nanopoulos,
Phys.\ Lett.\ B \textbf{257} (1991), 278-284
doi:10.1016/0370-2693(91)91893-Z

\bibitem{Kanti:1995vq}
P.~Kanti, N.~Mavromatos, J.~Rizos, K.~Tamvakis and E.~Winstanley,
Phys. Rev. D \textbf{54} (1996), 5049-5058
doi:10.1103/PhysRevD.54.5049 [arXiv:hep-th/9511071 [hep-th]].

\bibitem{Kanti:1997br}
P.~Kanti, N.~Mavromatos, J.~Rizos, K.~Tamvakis and E.~Winstanley,
Phys. Rev. D \textbf{57} (1998), 6255-6264
doi:10.1103/PhysRevD.57.6255 [arXiv:hep-th/9703192 [hep-th]].

\bibitem{Easson:2020mpq}
D.~A.~Easson, T.~Manton and A.~Svesko,
JCAP \textbf{10} (2020), 026 doi:10.1088/1475-7516/2020/10/026
[arXiv:2005.12292 [hep-th]].

\bibitem{Rashidi:2020wwg}
N.~Rashidi and K.~Nozari,
Astrophys. J. \textbf{890}, 58
doi:10.3847/1538-4357/ab6a10
[arXiv:2001.07012 [astro-ph.CO]].

\bibitem{Odintsov:2023aaw}
S.~D.~Odintsov, V.~K.~Oikonomou and F.~P.~Fronimos,
Phys. Rev. D \textbf{107} (2023), 08
doi:10.1103/PhysRevD.107.084007
[arXiv:2303.14594 [gr-qc]].

\bibitem{Odintsov:2023lbb}
S.~D.~Odintsov and T.~Paul,
Phys. Dark Univ. \textbf{42} (2023), 101263
doi:10.1016/j.dark.2023.101263
[arXiv:2305.19110 [gr-qc]].

\bibitem{Odintsov:2023weg}
S.~D.~Odintsov, V.~K.~Oikonomou, I.~Giannakoudi, F.~P.~Fronimos and E.~C.~Lymperiadou,
[arXiv:2307.16308 [gr-qc]].


\bibitem{Oikonomou:2022xoq}
V.~K.~Oikonomou,
Astropart. Phys. \textbf{141} (2022), 102718
doi:10.1016/j.astropartphys.2022.102718 [arXiv:2204.06304
[gr-qc]].

\bibitem{Nojiri:2023jtf}
S.~Nojiri and S.~D.~Odintsov,
[arXiv:2308.06731 [gr-qc]].

\bibitem{TerenteDiaz:2023kgc}
J.~J.~Terente D\'\i{}az, K.~Dimopoulos, M.~Kar\v{c}iauskas and
A.~Racioppi,
[arXiv:2310.08128 [gr-qc]].

\bibitem{Kawai:2023nqs}
S.~Kawai and J.~Kim,
[arXiv:2308.13272 [astro-ph.CO]].




\bibitem{Kawai:2021edk}
S.~Kawai and J.~Kim,
Phys. Rev. D \textbf{104} (2021) no.8, 083545
doi:10.1103/PhysRevD.104.083545 [arXiv:2108.01340 [astro-ph.CO]].



\bibitem{Kawai:2017kqt}
S.~Kawai and J.~Kim,
Phys. Lett. B \textbf{789} (2019), 145-149
doi:10.1016/j.physletb.2018.12.019 [arXiv:1702.07689 [hep-th]].


\bibitem{Choudhury:2023kam}
S.~Choudhury,
[arXiv:2307.03249 [astro-ph.CO]].








\bibitem{TheLIGOScientific:2017qsa}
B.~P.~Abbott \textit{et al.} [LIGO Scientific and Virgo],
Phys. Rev. Lett. \textbf{119} (2017) no.16, 161101
doi:10.1103/PhysRevLett.119.161101 [arXiv:1710.05832 [gr-qc]].

\bibitem{Monitor:2017mdv}
B.~P.~Abbott \textit{et al.} [LIGO Scientific, Virgo, Fermi-GBM
and INTEGRAL],
Astrophys. J. Lett. \textbf{848} (2017) no.2, L13
doi:10.3847/2041-8213/aa920c [arXiv:1710.05834 [astro-ph.HE]].

\bibitem{GBM:2017lvd}
 B.~P.~Abbott {\it et al.}
 ``Multi-messenger Observations of a Binary Neutron Star Merger,''
 Astrophys.\ J.\ {\bf 848} (2017) no.2, L12
 doi:10.3847/2041-8213/aa91c9
 [arXiv:1710.05833 [astro-ph.HE]].

\bibitem{Odintsov:2020sqy}
S.~D.~Odintsov, V.~K.~Oikonomou and F.~P.~Fronimos,
Nucl. Phys. B \textbf{958} (2020), 115135
doi:10.1016/j.nuclphysb.2020.115135 [arXiv:2003.13724 [gr-qc]].

\bibitem{Oikonomou:2021kql}
V.~K.~Oikonomou,
Class. Quant. Grav. \textbf{38} (2021) no.19, 195025
doi:10.1088/1361-6382/ac2168 [arXiv:2108.10460 [gr-qc]].

\bibitem{Oikonomou:2022ksx}
V.~K.~Oikonomou, P.~D.~Katzanis and I.~C.~Papadimitriou,
Class. Quant. Grav. \textbf{39} (2022) no.9, 095008
doi:10.1088/1361-6382/ac5eba [arXiv:2203.09867 [gr-qc]].

\bibitem{Anson:2019uto}
T.~Anson, E.~Babichev, C.~Charmousis and S.~Ramazanov,
JCAP \textbf{06} (2019), 023
doi:10.1088/1475-7516/2019/06/023
[arXiv:1903.02399 [gr-qc]].

\bibitem{Dalang:2020eaj}
C.~Dalang, P.~Fleury and L.~Lombriser,
Phys. Rev. D \textbf{103} (2021) no.6, 064075
doi:10.1103/PhysRevD.103.064075
[arXiv:2009.11827 [gr-qc]].



{



}










\bibitem{deHaro:2023lbq}
J.~de Haro, S.~Nojiri, S.~D.~Odintsov, V.~K.~Oikonomou and S.~Pan,
Phys. Rept. \textbf{1034} (2023), 1-114
doi:10.1016/j.physrep.2023.09.003 [arXiv:2309.07465 [gr-qc]].



\end{thebibliography}
\end{document}